\documentclass[11pt]{article}
\usepackage{amssymb,cite,graphicx}
\newcommand\blank[1]{}

\newcommand\toline[1]{--#1}

\newcommand{\fract}[2]{{\textstyle\frac{#1}{#2}}}

\newcommand{\ri}{\right}

\newcommand{\lf}{\left}
\newcommand{\te}{\theta}

\newcommand{\D}{{\rm D}}
\newcommand\ZZ{{\mathbb Z}}

\newcommand\NN{{\mathbb N}}
\newcommand\TT{{\mathbb T}}
\newcommand\AAa{{\mathbb A}}
\newcommand\HH{{\mathbb H}}
\newcommand\LL{{\mathbb L}}
\newcommand\MM{{\mathbb M}}

\newcommand{\balpha}{\alpha\kern -6.7pt\alpha}
\newcommand{\bbalpha}{\alpha\kern -4.95pt\alpha}
\newcommand\phup{^{\phantom p}}

\newcommand{\CaC}{{\cal C}}
\newcommand{\CH}{{\cal H}}

\newcommand{\CQ}{{\cal Q}}
\newcommand{\CQJ}{{\cal Q}\phup_J}
\newcommand{\bg}{{\bf g}}

\newcommand\eq{\begin{equation}}
\newcommand\en{\end{equation}}
\newcommand\bea{\begin{eqnarray}}
\newcommand\eea{\end{eqnarray}}
\newcommand\nn{\nonumber}

\newcommand{\wt}{\widetilde}

\newcommand{\One}{{\hbox{{\rm 1{\hbox to 1.5pt{\hss\rm1}}}}}}
\renewcommand{\One}{{\mathbb 1}}
\renewcommand{\One}{{\rm 1\!\!1}}

\newcommand{\resection}[1]{\setcounter{equation}{0}\section{#1}}

\begin{document}
\begin{titlepage}
\vskip 0.5cm
\begin{flushright}
DCTP/01/19  \\
{\tt hep-th/0103051}\\
\end{flushright}
\vskip 1.8cm
\begin{center}
{\Large \bf Spectral equivalences, Bethe Ansatz equations,\\[3pt]
 and reality properties in $\cal {P}\cal{T}$-symmetric \\[7pt] 
quantum mechanics} 
\end{center}
\vskip 0.8cm
\centerline{Patrick Dorey%
\footnote{\tt p.e.dorey@durham.ac.uk},
Clare Dunning%
\footnote{\tt tcd1@york.ac.uk}
and Roberto Tateo%
\footnote{\tt roberto.tateo@durham.ac.uk}
}
\vskip 0.9cm
\centerline{${}^{1,3}$\sl\small Dept.~of Mathematical Sciences,
University of Durham, Durham DH1 3LE, UK\,}
\vskip 0.2cm
\centerline{${}^2$\sl\small 
Dept.~of Mathematics, University of York, York YO10 5DD, UK }
\vskip 1.25cm
\vskip 0.9cm
\begin{abstract}
\vskip0.15cm
\noindent
The one-dimensional Schr\"odinger equation for the potential 
$x^6+\alpha x^2 +l(l+1)/x^2$  has many interesting properties. 
For certain  values of the parameters $l$ and $\alpha$ the equation is
in turn supersymmetric (Witten), quasi-exactly
solvable (Turbiner), 
and it also appears in Lipatov's approach to high energy QCD. 
In this  paper we signal some further  curious features 
of these theories, namely  novel  spectral equivalences with 
particular  second- and third-order differential equations.
These relationships are obtained
via a recently-observed connection between the theories of ordinary
differential equations and integrable models.
Generalised supersymmetry transformations acting
at the  quasi-exactly solvable points are 
also pointed out, and an efficient numerical procedure for the
study of these and related problems is described.
Finally we generalise slightly and then prove
a conjecture due to
Bessis, Zinn-Justin, Bender and Boettcher,
concerning the reality of the spectra of
certain ${\cal PT}$-symmetric quantum-mechanical systems.

\end{abstract}
\end{titlepage}
\setcounter{footnote}{0}
\def\thefootnote{\fnsymbol{footnote}}
\resection{Introduction}
The main subject of this paper will be the spectrum of Schr\"odinger equation
\eq
\CH(\alpha,l)\,\psi(x)=
\Bigl[-\frac{d^2}{dx^2}+x^{6}+\alpha x^{2}+ \frac{l(l+1)}{x^2} 
\Bigr]\psi(x)=E\,\psi(x)\,,
\label{x6}
\en
with boundary conditions $\psi(x)\to 0$ as $x\to\infty$ and 
$\psi(x) \sim x^{l+1}$  as $x\to 0$.\footnote{ 
The other natural boundary condition at the origin is
$\psi(x)\sim x^{-l}$; when we are interested in this spectral problem,
we shall write the differential operator as
$\CH(\alpha,-1{-}l)$.}

Over the years, many interesting features of this problem have been uncovered.
For $\alpha<0$ and $l(l{+}1)=0$, the theory corresponds to a double-well potential
on the full real line, with $l=-1$ and $l=0$ selecting the even and odd 
wavefunctions respectively. Such potentials have long been
studied as toy models for instanton effects  in  
quantum field theories~\cite{Instref}.
Furthermore, at $\{\alpha=-3,l=-1 \}$ 
the  
ground-state energy $E_0$ is zero, with the 
remaining energy levels  matching
those at $\{\alpha=3,l=0 \}$. This reflects the fact that
for these values of the parameters
the  model provides one of the simplest examples of
supersymmetric quantum mechanics~\cite{Wit}.
More peculiar are the properties of the spectrum at
 $\alpha=-(4J+2l+1)$: a finite number of energy 
levels can  be exactly computed~\cite{Tur,Ush}
being  roots of particular polynomials~\cite{BD}.
At these points, the model is said to be `quasi-exactly solvable'.
Finally, at least a couple  of physically 
interesting  spectral problems
are related to the above equation via simple variable and gauge
transformations.
The first is a linear combination of the harmonic and Coulomb 
potentials:
\eq
\Bigl[-\frac{d^2}{dx^2}+x^{2}-\frac{\sigma}{x}+ \frac{\gamma(\gamma+1)}{x^2} 
\Bigr]
\varphi(x)=\Lambda \;\varphi(x) \, , 
\label{spectii}
\en
with the two sets of parameters being related by
$\gamma=(l{-}1/2)/2,\;
\Lambda= -\alpha/2,\; 
\sigma= 2^{-3/2} E$.
The second theory is
\eq
\Bigl[-\frac{d^2}{dx^2}+\frac{1}{x^{3/2}}+\frac{\beta}{x}+ 
\frac{\delta(\delta+1)}{x^2} \Bigr] 
\chi(x)= \Delta \; \chi(x)\,,
\en 
with $\beta=16\alpha E^{-2},\; \delta=(l{-}3/2)/4,\; \Delta= -4096 E^{-4}$. 
Surprisingly,  at $\beta=0$ and $\delta(\delta{+}1)=0$ 	the latter  
equation is related to the Odderon problem in QCD~\cite{Lip}.

In this paper we discuss some further exact spectral relationships,
relating the problems (\ref{x6})  at different values of $\alpha$ and
$l$, and linking them to certain third-order 
differential  equations. We shall find these equivalences in 
the framework of the `ODE/IM correspondence'~\cite{DTa}. 
Some features of this correspondence are reviewed in sections 2
and 3, and they are then used to establish five spectral equivalences in
sections 4, 5, 6 and 7. Section 8 provides an alternative insight into the
second and third
equivalences using the so-called Bender-Dunne
polynomials, and in the process uncovers differential operators which
generalise the action of the supersymmetry generators to all quasi-exactly
solvable points. The conclusions are in section 9, and there are
two relatively self-contained appendices. The first explains a simple but
efficient
numerical approach to Schr\"odinger problems with polynomial potentials,
and the second uses ideas associated with the ODE/IM correspondence
to prove a generalised version of a conjecture due to
Bessis, Zinn-Justin, Bender and Boettcher.
\smallskip
\resection{Bethe ansatz equations for the $x^{2M}+x^{M-1}$ potential}
The r\^ole of functional relations in the spectral theory of
the Schr\"odinger equation has been extensively explored by
Voros~\cite{voros}, but only recently has it been
realised that they can lead, 
in certain cases, to a precise connection with the theory of
integrable models~\cite{DTa}. This so-called `ODE/IM correspondence' has
been developed in a number of
directions~\cite{BLZa,Sa,DTb,DTc,Sb,Sc,DDT2,Suzatt}, some of which are
reviewed in \cite{DDTr}.
In this section, we summarise the  results obtained 
in this context by Suzuki~\cite{Sc} concerning the Schr\"odinger equation
with potential $x^{2M}+\alpha x^{M{-}1}$, which includes the sextic potential
(\ref{x6}) as a special case.
Our treatment is perhaps a marginal simplification
of that given in \cite{Sc}, but the only genuinely new
contribution to the discussion is the inclusion of the angular 
momentum term. 
The  ODE to be considered is 
\eq
\CH(M,\alpha,l)\, \Phi(x)= 
\Bigl[-\frac{d^2}{dx^2}+x^{2M}+\alpha x^{M-1}+ 
\frac{l(l+1)}{x^2} \Bigr]\Phi(x)=E\,\Phi(x)
\label{sh}
\en
with $M$ a positive real number, which for technical reasons will
sometimes be taken to be greater than $1$.
The goal is to find the eigenvalues $\{ E_i \}$, those $E$ for 
which 
(\ref{sh}) has a solution vanishing 
as $x\to +\infty$, and behaving as $x^{l+1}$ as $x\to 0$.
(For $\Re e\,l>-1/2$, the latter condition is equivalent to the demand that
the usually-dominant $x^{-l}$ behaviour near the origin should be absent; 
outside this region, the problem is best
defined by analytic 
continuation\footnote{For a discussion, see chapter~4 of \cite{NEWT}.}.)
The starting-point is
the uniquely-determined solution 
$Y(x,E,\alpha,l)$ which has the following asymptotic for large,
positive  $x$\,\cite{Sib}\,:
\eq
Y(x,E,\alpha,l) \sim 
{ 
x^{-M/2 -\alpha/2} 
\over 
\sqrt{2 i} 
}  
\exp \lf( - \fract{x^{M+1}}{M+1} \ri) ~,
\en
and
an associated set of functions $Y_k$\,:
\eq
Y_k(x,E,\alpha,l)= \Omega^{k/2 +  k\alpha/2}
Y(\Omega^{-k}  x,\;\Omega^{-2 M k} E,\; \Omega^{(M+1)k}\alpha,\;l)
~,\quad
\Omega= \exp( \fract{i \pi}{M+1})~.
\en
For integer $k$ the $Y_k$'s are solutions of~(\ref{sh}),
and any pair  $\{Y_k,Y_{k+1} \}$ forms a basis of solutions. 
In particular, $Y_{-1}$ can be written as a  linear combination of 
$Y_0$ and $Y_{1}$, 
the precise relation being
\eq
T(E,\alpha,l) Y_0(x,E,\alpha,l)=Y_{-1}(x,E,\alpha,l)+
Y_{\;1}(x,E,\alpha,l)\,.
\label{Ty}
\en
The coefficient 
\eq
T(E,\alpha,l)=W[\,Y_{-1}\,, Y_1]\,
=
\mbox{Det}
\lf[\matrix{Y_{-1}(x)  & Y_{1}(x)  \cr
Y'_{-1}(x) & Y'_{1}(x)}\ri]\,,
\label{tform}
\en
given here as a Wronskian,
is called a Stokes multiplier.
In terms of the original function $Y(x,E,\alpha,l)$, 
(\ref{Ty}) taken at $\alpha$ and at $-\alpha$
leads to the following pair of equations:
\eq
T^{(+)}(E) Y^{(+)}(x,E) =\Omega^{-(1+\alpha)/2}  
Y^{(-)}(\Omega  x, \Omega^{2 M}  E)+
 \Omega^{(1+\alpha)/2} Y^{(-)}(\Omega^{-1}  x,\Omega^{-2 M}  E)~
\label{peqpa}
\en
\eq
T^{(-)}(E) Y^{(-)}(x,E) =\Omega^{-(1-\alpha)/2}  
Y^{(+)}(\Omega  x, \Omega^{2 M}  E)+
 \Omega^{(1-\alpha)/2} Y^{(+)}(\Omega^{-1}  x,\Omega^{-2 M}  E) ~
\label{peqp}
\en
where
\eq
T^{(\pm)}(E) = T(E, \pm \alpha,l)\, ,~~~~Y^{(\pm)}(x,E)=
Y(x,E,\pm \alpha,l)\, .
\en
Keeping $\Re e\,l>-1/2$, the leading behaviour of $Y$ near the origin at generic $E$ is 
\eq
Y(x,E,\alpha,l) \sim D(E,\alpha,l ) x^{-l}+\dots~.
\label{Ynear0}
\en
At the zeroes of $D(E)$, 
the leading term is instead proportional to
$x^{l+1}$, and $Y(x,E,\alpha,l)$ is an eigenfunction of (\ref{sh}).
This implies that $D(E)$ is proportional to the spectral determinant for 
the problem. (For $M>1$
the order of $D(E)$ can be shown to be less than one, so $D(E)$ 
is fixed up to a constant by the positions of its zeroes.) 
Setting   $D^{(\pm)}(E)=D(E,\pm\alpha,l)$, from (\ref{peqpa}) and (\ref{peqp}) 
we have
\bea
T^{(+)}(E) D^{(+)}(E) &=&\Omega^{-(2l+1+\alpha)/2}  
D^{(-)}(\Omega^{2M} E)+
\Omega^{(2l+1+\alpha)/2} D^{(-)}(\Omega^{-2M}  E)\, ,
\label{tq}\\[4pt]
T^{(-)}(E) D^{(-)}(E) &=&\Omega^{-(2l+1-\alpha)/2}  
D^{(+)}(\Omega^{2M} E)+
\Omega^{(2l+1-\alpha)/2} D^{(+)}(\Omega^{-2M}  E)\, . \qquad
\quad
\label{tqa}
\eea
Now let the  zeroes of $D^{(\pm)}(E)$ be at
$\{ E^{(\pm)}_k \}$, and set $E=E^{(\pm)}_k$ in either (\ref{tq}) or
(\ref{tqa}).  
Both $T^{(\pm)}(E)$ and $D^{(\pm)}(E)$ are  entire in $E$, so the LHS  
of the relevant equation vanishes. Factorising the functions $D^{(\pm)}(E)$ 
as products over their zeroes, for $M>1$
the following system of `Bethe ansatz' type
equations for the energy spectrum is obtained:
\bea
\prod_{n=0}^{\infty} \lf( { E^{(-)}_n - \Omega^{-2M} \, E^{(+)}_k  \over 
E^{(-)}_n -  \Omega^{\;2M} \, E^{(+)}_k}
\ri)
&=& - \Omega^{-2l-1 - \alpha } \, ; \label{ba01} \\
\prod_{n=0}^{\infty} \lf( { E^{(+)}_n - \Omega^{-2M} \, 
E^{(-)}_k   \over 
E^{(+)}_n - \Omega^{\;2M} \, E^{(-)}_k }
\ri)
&=& - \Omega^{-2l-1 + \alpha }\, . 
\label{ba02}
\eea
Note that  the spectra of the  Hamiltonians 
 $\CH(M,\alpha,l)$ and $\CH(M,-\alpha,l)$ are  completely tangled  up
by the Bethe ansatz constraints.

One of the products of the ODE/IM correspondence was the realisation that
energy levels for Schr\"odinger problems can be calculated using 
nonlinear integral equations~\cite{DTa}.  
For the ODE (\ref{sh}) with $l{=}0$ and 
$\alpha$ small, these were
derived in \cite{Sc}, and it is straightforward to include the effect of
the angular momentum term. Suppose that\\
 {\bf (a)} all the zeroes of
$D^{(\pm)}(E)$ lie on the positive real axis of the complex $E$ plane, and\\
{\bf (b)} all the zeroes of $T^{(\pm)}(E)$ lie away from it.\\
(As shown
in appendix B, {\bf (a)} holds if $l>-1/2$, and {\bf (b)} 
if $|\alpha|<M+1-|2l{+}1|$.\,)
Setting
\eq
a^{\pm}(\theta) = \Omega^{ 2l+1 \pm \alpha} \frac{ D^{(\mp)} ( \Omega^{-2M} E) }
{ D^{(\mp)} ( \Omega^{2M} E) }
\qquad\mbox{with}\quad E=\exp\left(\frac{2M}{M{+}1}\,\theta\right)\,,
\en
the equations are then
\bea
\ln a^{\pm}(\te) &=& i \frac{\pi}{2} (2l+1\pm \frac{\alpha}{M}) 
- i b_0 \, e^{\te} \nn \\ 
&& \hspace {-1cm}
 + \int_{{\cal C}_1} d\te' \, K_1(\te-\te') \ln
(1+a^{\pm}(\te')) 
 - \int_{{ \cal C}_2} d\te' \, K_1(\te-\te') \ln
(1+\frac{1}{a^{\pm}(\te')})\qquad \nn \\
&& \hspace {-1cm}
 + \int_{{\cal C}_1} d\te' \, K_2(\te-\te') \ln
(1+a^{\mp}(\te')) 
 - \int_{{ \cal C}_2} d\te' \, K_2(\te-\te') \ln
(1+\frac{1}{a^{\mp}(\te')})\qquad
\label{nlies}
\eea
The integration contours ${\cal C}_1$ and ${\cal C}_2$ run just
below and just above the real axis respectively,
and the constant 
$b_0 = \pi^{1/2} \Gamma(\frac{1}{2M})/(2 M \,
\Gamma(\frac{3}{2}+\frac{1}{2M}))\,$ is fixed via a consideration of the WKB
asymptotics of $D^{(\pm)}(E)$ for $|E|\to\infty$, $\arg(E)\neq 0$.
An integral expression 
for the  kernels $K_1$ and $K_2$ at general values of $M$ 
was found in \cite{Sc}.
In this paper we are particularly interested in the sextic
potential, and here
we note that the special nature  of this case 
is reflected in the fact that at $M=3$ 
the kernel functions can be explicitly integrated, and  have the simple 
forms
\eq
K_1(\te)=- \frac{ \sqrt 3}{2\pi(2\cosh \te +1)}\,, \quad \quad
K_2(\te)=- \frac{ \sqrt 3}{2\pi(2\cosh \te -1)}~.
\en
A search along the real $\te$
axis for the zeroes of the functions
  $1+a^{\pm}(\te)$ provides the energy levels of the
Hamiltonians $\CH(M,\pm \alpha,l)$.
While we do not have a rigorous proof, we expect that the solution to
(\ref{nlies}), which can readily be obtained numerically by iteration, is {\em
unique} for $\alpha$ and $l$ in the stated range. This is one way to justify the 
claim that, for $l>-1/2$ and $|\alpha|<M+1-|2l{-}1|$, the Bethe Ansatz
equations (\ref{ba01}) and (\ref{ba02}), together with the `analytic
properties' {\bf (a)} and {\bf (b)} and the WKB asymptotic which determined
$b_0$, characterise the set of numbers $\{E^{(\pm)}_k\}$ uniquely.
(An alternative approach to this question might be to generalise the analysis of
\cite{DTa,BLZa} and appendix~A of \cite{BLZneq}, based on the so-called 
quantum Wronskian relations.)

To treat more general values of $\alpha$ and $l$, 
we will find it
most convenient to work directly with the Bethe ansatz equations.
In contrast to the integral equation, these do not have a unique solution.
As is standard in studies of the Bethe ansatz, it is useful to take logarithms.
For $l>-1/2$ and 
$|\alpha| < M+2l+2$ (see appendix~\ref{appbb}), all 
energies  
$E_k^{(\pm)}=E_0^{(\pm)},E_1^{(\pm)},E_2^{(\pm)} \dots$ 
are positive. For this situation we assume that
there  is a one-to-one correspondence
between  these
energies  
and the  integers $k=0,1,2, \ \dots$ :
\bea 
\sum_{n=0}^{\infty}
\ln \lf(  { E_n^{(-)} -\Omega^{-2M}  \; E_k^{(+)} 
\over E_n^{(-)} - \Omega^{2M} \; E_k^{(+)}}\ri)
&=& - i \pi \left[  \frac{2l+1 + \alpha}{M+1 } +  2 k +1 \right] \, ;
\label{baloga}\\
\sum_{n=0}^{\infty}
\ln \lf(  { E_n^{(+)} -\Omega^{-2M}  \; E_k^{(-)} 
\over E_n^{(+)} - \Omega^{2M} \; E_k^{(-)}}\ri)
&=& - i \pi \left[ \frac{2l+1 - \alpha}{M+1 } +  2 k +1 \right]  \, ,
\label{balogb}
\eea
where the logarithms are all on the principal branch: 
$-\pi \le -i \ln < \pi$.
At larger
values of
$|\alpha|$ and $|l|$ some of the low-lying energies might become negative,
and in such cases
care must be taken to keep track of the nontrivial monodromy of the 
log function.
%
%
\smallskip
\resection{The Bethe ansatz approach to a third-order equation}
This section summarises 
the derivation of equations of Bethe ansatz type for a third-order ODE with
`potential' $x^{3N}$, following \cite{DTc,DDT2}.
We start with the equation
\eq 
\Bigl[\,\frac{d^3}{dx^3}+x^{3N}+\frac{L}{x^3}-G \Bigl( 
 \frac{1}{x^2}\frac{d}{dx}- \frac{1}{x^3} \Bigr)\,\Bigr]\phi(x)= E \; \phi(x)\,, 
\label{tsh}
\en
and, as in~\cite{DDT2}, 
rewrite it as 
\eq
\lf [\,\D({\bf g})+x^{3N}\, \ri]\phi(x) = E \;  \phi(x) \,  
\label{invH}
\en
where 
$\bg= \{g_0,g_1,g_2 \}$ with $g_0+g_1+g_2=3$, and
\eq
\D({\bf g})=\D(g_2-2)\D(g_{1}-1)\D(g_0)~,\quad
\D(g)= \left( \frac{d}{dx} - \frac{g}{x} \right)~.
\en
The relationship between ${\bf g}$ and 
$\{G,L \} $ is
\eq
G= g_0 g_1 + g_0 g_2 + g_1 g_2 -2~~~~,~~~~
L=2-g_0 g_1 g_2- (g_0 g_1 + g_0 g_2+ g_1 g_2)\, . 
\en 
Again we introduce  a uniquely-defined   function 
$y(x,E,\bg)$, which solves  
(\ref{tsh}) and tends to 
zero   as $x\to\infty$ along the positive real axis as
\eq
y(x, E, \bg)  \sim \frac{x^{-N}\!}{i \sqrt{3}} 
\exp \lf( - \fract{x^{N+1}}{N+1} \ri)\, . 
\en 
Given $y(x)$, bases of solutions 
are constructed just as in the 
second-order case. Set 
\eq
y_k(x,E,\bg)= 
\omega^k y(\omega^{-k} x, \omega^{-3 N k } \,E, \bg)\, ,~~~~~
\omega= \exp \lf( \fract{2\pi i}{3 N+3}\ri)\, . 
\label{ydef}
\en
For integer $k$, $y_k$ solves (\ref{tsh}), and 
$\{y_k,y_{k+1},y_{k+2}\}$ form a basis.
We can therefore expand $y_0$ as
\eq
y_0 - C^{(1)} y_1 +  C^{(2)} y_2 -  y_3=0
\en
with coefficients  $C^{(1)}$ and $C^{(2)}$ -- Stokes
multipliers -- which
are independent of $x$.  Eliminating $C^{(2)}$, we have
\eq
W_{02} - C^{(1)} W_{12} + W_{23}=0 \, ,
\en
where the Wronskians of pairs of solutions,
\eq
W_{k_1\, k_2}= W[y_{k_1}\; y_{k_2}]= 
\mbox{Det}
\lf[\matrix{y_{k_1}(x)  & y_{k_2}(x)  \cr
y'_{k_1}(x) & y'_{k_2}(x)          }\ri]
\label{y=det} \, ,
\en
were used.
(Note, since $y_{k_1}$ and $y_{k_2}$ solve a {\em third}-order equation,
these are nontrivial functions of $x$.)
Now multiply by $y_1$ and use the relation $y_1 W_{02}=y_0 W_{12} +y_2 W_{01}$
to find 
\eq
 C^{(1)} y_1 W_{12} =y_0 W_{12} +y_2 W_{01} + y_1 W_{23} \, .
\label{cyw1}
\en 
In this form the relation can be rewritten in terms of just two 
functions, $y(x,E,{\bf g})$ and
$W(x,E,\bg)=W_{01}(x,E, \bg)$ :
\bea
 C^{(1)}(E) y( \omega^{-1} x,\omega^{-3N} E ) W( \omega^{-1} x, \omega^{-3N} E )=
\omega^{-1} y(x,E) W( \omega^{-1} x,  \omega^{-3N} E) 
&+& \nn \\
 y(\omega^{-2} x, \omega^{-6N} E) W(x, E) + 
 \omega y( \omega^{-1} x ,  \omega^{-3N} E ) 
W(\omega^{-2} x , \omega^{-6N}E)\, . 
&{}& 
\label{cyw}
\eea
We initially suppose that
$\Re e(g_0) < \Re e(g_1) <\Re e(g_2)$. Then as 
$x \rightarrow 0$ the leading  behaviours of $y$ and $W$ are
\eq
y(x) \sim D^{(1)}(E,\bg)\, x^{g_0}\, ,~~~~~W(x) \sim 
D^{(2)}(E,\bg)\, x^{g_0+g_1-1}\, .
\label{YWnear0}
\en
Using equations (\ref{YWnear0}) the relation  (\ref{cyw}) becomes 
\bea
C^{(1)}(E) 
D^{(1)}(\omega^{-3N} E )  D^{(2)}(\omega^{-3N} E )   = 
\omega^{g_0 -1} 
D^{(1)}(E ) D^{(2)}(\omega^{-3N}  E)  
 +~~~ \qquad&{}& \nn\\
 \omega^{g_1-1}
D^{(1)}( \omega^{-6N}  E ) D^{(2)}(  E) 
+ \omega^{2- g_0 -g_1}
D^{(1)}(\omega^{-3N}  E )  D^{(2)}(\omega^{-6N} E )\, . 
&{}& 
\label{cDD}
\eea
Again we shall consider this expression at the 
zeroes of $D^{(1)}$ and  $D^{(2)}$. It is convenient to write 
these as $E^{(1)}_k$ and $\omega^{3N/2}E^{(2)}_k$\,, so that
\eq
D^{(1)}(E_k^{(1)}, \bg)=0\, ,~~~~~D^{(2)}(\omega^{3 N/2} E_k^{(2)} , \bg )=0\, .
\en
As in the second-order case, these functions have a spectral interpretation.
In particular, the vanishing of $D^{(1)}(E)$ signals the existence of a
solution, at that value of $E$,
to (\ref{invH}), decaying as $x\to\infty$, and having a
faster-than-usual decay at the origin:
\eq
y(x) \sim x^{\min(g_1,g_2)}\,,\qquad
x\to 0 \,.
\en
(Since $g_1$ and $g_2$ can be complex, by $\min(g_1,g_2)$ we mean
whichever of $g_1$ and $g_2$ has the smallest real part.)
Evaluating (\ref{cDD}) at $E \in \{ E_k^{(1)}\}$ and $E \in \{
\omega^{3N/2}E_k^{(2)}\}$			 
and imposing the  entirety of $C^{(1)}(E)$ leads to the following set of
$SU(3)$-related BA 
equations, with $k=0,1,2,\dots$\,:
\bea
\prod_{j=0}^\infty
\left(
\frac{ E_{j}^{(1)} - \omega^{-3N} \; E_k^{(1)} }
{ E_j^{(1)} - \omega^{3N} \; E_k^{(1)} } 
\right)
\left(
\frac{ E_j^{(2)} -  \omega^{\frac{3N}{2}} \; E_k^{(1)} } 
{ E_{j}^{(2)} - \omega^{-\frac{3N}{2}} \; E_k^{(1)}}
\right)
&=& 
- \omega^{g_0 - g_1}\, ;
\label{Beq1}
\\ 
\prod_{j=0}^\infty
\left(
\frac{ E_{j}^{(2)} -  \omega^{-3N} \; E_k^{(2)}
}
{ E_{j}^{(2)} - \omega^{3N} \; E_k^{(2)}} 
\right)
\left(
\frac{ E_{j}^{(1)} -  \omega^{\frac{3N}{2}} \; E_k^{(2)} }
{ E_{j}^{(1)} -  \omega^{-\frac{3N}{2}} \; E_k^{(2)}} 
\right)
&=& 
- \omega^{2 g_1 + g_0 - 3}\,.
\label{Beq2}
\eea
Since $g_0{+}g_1{+}g_2=3$, the right-hand sides of 
equations (\ref{Beq1}) and (\ref{Beq2}) can be given a
more symmetrical appearance by rewriting them as
$-\omega^{2 g_0 +g_2 -3}$ and $-\omega^{-2 g_2 - g_0 +3}$ respectively.
These equations, together with WKB-like asymptotics for 
$D^{(1)}(E)$ and $D^{(2)}(E)$, fix
 the numbers $E_k^{(1)}$ and  $E_k^{(2)}$
up to discrete ambiguities, which for the problems in hand can be eliminated
given some facts about the approximate positions of the zeroes of the
functions $D^{(1,2)}(E)$ and some associated functions $a^{(1,2)}(E)$. These
are analogous to the analyticity conditions ${\bf (a)}$ and ${\bf (b)}$ of the
previous section, and are described in more detail in
\cite{DTc,DDT2}. It is also possible to solve the system via
a nonlinear integral equation, but this will not be needed here.
\smallskip
\resection{The first spectral equivalence}
The first spectral equivalence follows from observing 
that at $N=1$,  $\omega^{3N}=-1$ and  
$\omega^{\frac{3N}{2}}=i$. The $SU(3)$ BA equations therefore simplify
to
\bea
\prod_{j=0}^\infty
\left(
\frac{ E_{j}^{(2)} - i E_k^{(1)}}
{ E_{j}^{(2)} +i  E_k^{(1)}} 
\right)
&=& 
-  \omega^{2 g_0 +g_2 -3} \, ; \label{finalba1}\\
\prod_{j=0}^\infty
\left(
\frac{ E_{j}^{(1)} - i E_k^{(2)}}
{ E_{j}^{(1)} + i   E_k^{(2)}} 
\right)
&=& 
- \omega^{ -2 g_2 - g_0 +3}\, .
\label{finalba2}
\eea
These equations coincide with the system~(\ref{ba01}), (\ref{ba02}) at $M=3$ 
provided the right-hand sides of the two BA sets are  equated:
\eq
(2 g_0 +g_2 -3)/3=(-2l-1-\alpha)/4\, ,~~-(2 g_2 +g_0 -3)/3=(-2l-1+\alpha)/4 
\, .
\label{baeq}
\en
Combined with a matching of the analytic properties {\bf (a)} and {\bf (b)},
this suggests the following relationship 
between quantities in the two problems:
\bea
D^{(1)}(\kappa^{-1} E,\bg) &=& f(\alpha,l)\, D^{(+)}( E, \alpha, l)\, ,
\label{prop}
\\[3pt]
D^{(2)}( i \kappa^{-1}  E,\bg) &=& f(\alpha,l)\, D^{(-)}(E, \alpha, l)\, .
\eea
The proportionality factors $f(\alpha,l)$ 
and $\kappa$ cannot be determined by a comparison of
the Bethe ansatz equations alone. However, 
as in~\cite{DTc}, $\kappa$ can be calculated by comparing
the large negative $E$ asymptotics of $D^{(1)}$ and $D^{(2)}$. The
result, independent of 
$\alpha$ and $l$, is $\kappa=4/(3\sqrt 3)$.
Solving (\ref{baeq}), the parameters $(\alpha,l)$ and $\bg$ are related as
\eq
\alpha \equiv \alpha(g_0,g_2)=2(2-g_0-g_2)\, ,~~~l \equiv 
l(g_0,g_2)= (2 g_2 -3 -2 g_0)/6 \, , 
\label{alphag}
\en
and
\eq
g_0=(1 - \alpha - 6l)/4\, ,~~~ g_1=(1 + \alpha/2)~~,~~g_2=(7 - \alpha + 6l)/4 \, .
\label{galpha}
\en
Thus we have a  spectral equivalence  between the following 
eigenvalue problems
\eq 
\Bigl[-\frac{d^2}{dx^2}+x^{6}+\alpha x^{2}+ \frac{l(l+1)}{x^2} 
\Bigr]\Phi(x)=E\;\Phi(x)\, ,~~~~\Phi|_{x \rightarrow 0} \sim x^{l+1} \, ;
\label{sh2}
\en
\eq 
\kappa \Bigl[\frac{d^3}{dx^3}+x^{3}+\frac{L}{x^3}-G \Bigl( 
 \frac{1}{x^2}\frac{d}{dx}- \frac{1}{x^3} \Bigr)\Bigr]\phi(x)=  E \; \phi(x)
\, ,~~~~\phi|_{x \rightarrow 0} \sim x^{\min(g_1,g_2)}\, ,
\label{tsh2}
\en
where
\eq
G= g_0 g_1 + g_0 g_2 + g_1 g_2 -2\, ,~
L=2-g_0 g_1 g_2- (g_0 g_1 + g_0 g_2+ g_1 g_2)\, ,
\en 
and the parameters in the two models are related 
as in  (\ref{alphag}) and (\ref{galpha}).
Note that the general $SU(3)$-related equation at $N=1$ is mapped onto the
general sextic-potential problem. The number of parameters matches up, because
the third-order equation allows for two linearly-independent angular
momentum type terms. The different r\^oles that these parameters play
in the second-order problem will be important for the next spectral
equivalence that we discuss.
\smallskip
\resection{The second spectral equivalence}
Our second spectral equivalence is related to the  
enhanced symmetries of the third-order equation.
To be more precise, the differential equation
(\ref{invH}) is unchanged under permutations of $\{g_0,g_1,g_2\}$, while
the values of $\alpha$ and $l$ which appear in the corresponding
second-order equation, given by (\ref{alphag}), are
not.
If we make a continuation in 
$\{g_0,g_1,g_2\}$ which swaps $g_1$ and $g_2$ while
leaving $g_0$ unchanged, then both the third-order equation itself, and the
specification  (\ref{YWnear0}) of $D^{(1)}$, are unchanged, and so 
\eq
D^{(1)}(\kappa^{-1}E,\bg) 
\rightarrow 
D^{(1)}(\kappa^{-1}E,\bg)\,.
\en
Using (\ref{prop}), this means that
\eq
D^{(+)}(E, \wt{\alpha},\wt{l} ) 
\,=\,
\frac{f(\alpha,l)}{f(\wt\alpha,\wt l)}\,
D^{(+)}(E, \alpha, l) \,,
\label{spec2}
\en
with 
\eq
\wt{\alpha} \equiv \alpha(g_0,g_1)=(3 - \alpha + 6 l)/2 ~ , 
\quad
\wt{l} \equiv l(g_0,g_1)=(\alpha + 2 l -1)/4 \, .
\label{tl}
\en 
It will sometimes be convenient to put this in matrix form. If 
$\balpha= (\alpha,l,1)^T$, then
\eq
\wt\balpha=\TT\,\balpha~~,\quad \mbox{with}~~ \TT=
\left(\begin{array}{ccc} 
-1/2&3&3/2\\
1/4&1/2&-1/4\\
0&0&1
\end{array}\right)\,.
\en
At this stage we do not know how to calculate $f(\alpha,l)$ and
$f(\wt\alpha,\wt l)$ exactly, but
(\ref{spec2}) can be combined with (\ref{Dzero}) from 
appendix~B to give their ratio:
\eq
\frac{f\bigl(\,\alpha,l\,\bigr)}{f\bigl(\,\wt\alpha,\wt l\,\bigr)}\,=\,
\frac{\Gamma\bigl(\,\wt l+\frac{1}{2}\,\bigr)}
{\Gamma\bigl(\,l+\frac{1}{2}\,\bigr)}~.
\en
Note that this is singular or zero at negative-half-integer values of
$\tilde{l}$ or $l$, at which a `resonance' is expected in one or other of the
spectral problems \cite{NEWT}. Away from these points, we
have a  spectral equivalence between
\eq 
\Bigl[-\frac{d^2}{dx^2}+x^{6}+\alpha x^{2}+ \frac{l(l+1)}{x^2} 
\Bigr]\Phi(x)=E\;\Phi(x)~~~,~~~\Phi|_{x \rightarrow 0} \sim x^{l+1} \, , 
\label{sh3}
\en
and
\eq 
\Bigl[-\frac{d^2}{dx^2}+x^{6}+ \frac{(3 - \alpha + 6 l)}{2} x^{2}+ 
\frac{(\alpha + 2 l -1)(\alpha + 2l  +3)}{16 x^{2}} \Bigr]\Phi(x)=E\;\Phi(x) \, ,
\label{sh33}
\en
with $\Phi|_{x \rightarrow 0} \sim  x^{(\alpha + 2 l-1)/4+1}$.
An alternative viewpoint on this equivalence in terms of intertwining
operators will be given in \S8 below, while some direct
numerical checks are reported in appendix A.
\smallskip
\resection{The third spectral equivalence}
As was mentioned in the introduction, at the special values 
$\alpha=\alpha_J(l)=-(4J + 2 l+1)$, with $J$ a positive integer,
the model (\ref{x6}) is `quasi-exactly solvable' (QES), and
the first $J$  energy levels 
can be computed exactly.  
For $J=1$, the single exactly-solvable energy is the ground state 
and the model is an example of
supersymmetric quantum mechanics.
This is signalled by the fact that the Hamiltonian at 
$\alpha=\alpha_1(l)=-(2l{+}5)$ can factorised in terms of first-order operators as
\eq
{\CH}(\alpha_1,l)\equiv
\lf [-\frac{d^2}{dx^2}+x^{6}-(2l+5) x^{2}+ \frac{l(l+1)}{x^2} \ri]
=\CQ^- 
\CQ^+ \, ,
\en 
where
\eq
\CQ^- =
\lf [ -{d \over dx} +  x^3 - {l+1 \over x} \ri] 
\, ,~~~~
\CQ^+=
\lf [  {d \over dx} +  x^3 - {l+1 \over x} \ri]\, .
\en
The  SUSY `partner' Hamiltonian $\widehat {\CH}=
\CH(\widehat\alpha_1,\widehat l_1)$  
is obtained through the  intertwining  relation
$\CH(\widehat\alpha_1,\widehat l_1)\, \CQ^+ =\CQ^+ \, {\CH}(\alpha_1,l)$
with $\widehat\alpha_1(l)=1{-}2l$ and $\widehat l_1(l)=l{+}1$\,: 
\eq
\widehat{\CH}=  \CQ^+ 
\CQ^- =\lf [-\frac{d^2}{dx^2}+x^{6}+(1-2l) x^{2}+ \frac{(l+1)(l+2)}{x^2} 
\ri]
\, .
\en
The wavefunctions of the two models are simply related by
\eq
\widehat{\psi}_i(x) =  \CQ^+ \psi_i(x) \, .
\en
However,
the ground-state wavefunction of $\CH(\alpha,l)$ is
$ \psi_0= x^{l+1}\exp(-\fract{x^4}{4}) \, , $
and this is annihilated by $\CQ^+$.
As a result,
${\CH}$ and $\widehat{\CH}$ are spectrally equivalent 
save for the extra level at $E=0$ only present in  $\mbox{Spec}\, ({\CH})$. 
This (very standard) result makes it natural to ask whether similar
`partner potentials' might exist at higher values of $J$, sharing
the same spectra as the QES
Hamiltonians $\CH(\alpha_J(l),l)$
apart from the  first $J$ levels.
We shall find that this question has a surprisingly simple answer
by using the Bethe ansatz approach to the spectral problem.

Setting  $\alpha=\alpha_J(l)=-(4J {+}2l {+}1)$, the $J$ exactly-solvable 
levels 
$E^{(+)}_0\!$,
$E^{(+)}_1\!$, \dots
$E^{(+)}_{J-1}$
lie in the sector `$(+)$'. The BAE (\ref{baloga}), (\ref{balogb}) for $M{=}3$
are then
\bea
\sum_{n=0}^{\infty}
\ln
\lf( { E^{(-)}_n - i  \, 
E^{(+)}_k  \over 
E^{(-)}_n +i  \, E^{(+)}_k }
\ri)  
&=& -i \pi [-J +2 k +1]\, 
; \label{baa01} \\[3pt]
\hspace{-1cm}
\sum_{n=0}^{J-1}
\ln \lf( { E^{(+)}_n -i  \, E^{(-)}_k  \over 
E^{(+)}_n +i  \, E^{(-)}_k}
\ri)
+\sum_{n=J}^{\infty} \ln \lf( { E^{(+)}_n -i  \, E^{(-)}_k  \over 
E^{(+)}_n +i  \, E^{(-)}_k}
\ri) 
&=& -i \pi [l+3/2+J + 2 k] \,,  \label{baa02}
\eea
where the integer $k$ runs from 0 to $\infty$.
Next, we will use the fact that
the exactly-solvable energy levels appear symmetrically, as
$E_i^{(+)}=-E_{J-i-1}^{(+)}$, to simplify the first sum
on the LHS of 
(\ref{baa02}).
Recalling that, since $\alpha$ is negative,
the $E^{(-)}$ are positive for $l>-1/2$, and keeping track of 
the monodromy of the logarithms by using the reflection formula
\eq
\ln \lf ( { -x -i \over  -x + i} \ri) = -2\pi i- 
\ln \lf ( { x -i \over  x + i} \ri) \, ,~~~~~~(x \ge 0) \, ,
\en 
we obtain
\eq
\sum_{n=0}^{J-1} \ln  \lf({E^{(+)}_n -i\, E^{(-)}_k  \over 
E^{(+)}_n +i  \, E^{(-)}_k}
\ri) = -i \pi J \, .  
\en 
Finally, ignoring  the first $k=0,1,\dots J{-}1$ instances of (\ref{baa01}) and
 relabelling  $E^{(+)}_{k+J} \rightarrow  E^{(+)}_{k}$ we    end up with
\bea
\sum_{n=0}^{\infty} \ln\lf( { E^{(-)}_n - i  \, 
E^{(+)}_k   \over 
E^{(-)}_n +i  \, E^{(+)}_k }
\ri)
&=& -i \pi [J+2 k +1]  \, ;  \label{baa03} 
\\
\sum_{n=0}^{\infty} \ln\lf( { E^{(+)}_n -i  \, E^{(-)}_k  \over 
E^{(+)}_n +i  \, E^{(-)}_k}
\ri)
&=& -i \pi [l+2 k+3/2]  \, ,
\label{baa04}
\eea
where $k$ again runs from $0$ to $\infty$.
Comparing with (\ref{baloga}), (\ref{balogb}) 
we now reinterpret the left-hand sides
of equations (\ref{baa03}) and (\ref{baa04})
as the quantisation conditions for the energy levels of
a new potential, with parameters $\widehat{\alpha}_J$
and $\widehat l_J$\,:
\bea
-i\pi\bigl[J+2k+1\bigr] &=& 
-i\pi\left[\frac{2 \widehat{l}_J {+}1 {+} \widehat{\alpha}_J}{4}+2k+1\right]\,,
\\[2pt]
-i\pi\bigl[l+2k+3/2\bigr] &=& 
-i\pi\left[\frac{2 \widehat{l}_J {+}1 {-} \widehat{\alpha}_J}{4}+2k+1\right]\,.
\eea
Solving,
\eq
\widehat{\alpha}_J =2J - 2 l -1=-(3 + \alpha + 6 l)/2 ~ ,
\quad
\widehat{l}_J= J+l =(-\alpha + 2 l -1)/4 \, ,
\en
and this can again be put in matrix form, for $\balpha=(\alpha,l,1)^T$,
as
\eq
\widehat\balpha=\HH\,\balpha~~,\quad \mbox{with}~~ 
\HH= \left(\begin{array}{ccc} 
-1/2&-3&-3/2\\
-1/4&1/2&-1/4\\
0&0&1
\end{array}\right)\,.
\en
Then, for $J\in \NN$,
\eq
E^{(+)}_{k+J}(\alpha_J,l)=E^{(+)}_k(\widehat{\alpha}_J,\widehat{l}_J)
\, ,~~~(k=0,1,2,\dots) \, 
\en
and,
modulo the exactly-solvable levels, there is a  spectral equivalence
between
\eq 
\quad
\Bigl[-\frac{d^2}{dx^2}+x^{6} - (4 J {+} 2 l {+}1) x^{2}+ 
\frac{l(l{+}1)}{x^{2}} \Bigr]\Phi(x)=E\,\Phi(x)\, ,
\quad\qquad\Phi|_{x \rightarrow 0} \sim x^{l+1} \, 
\qquad
\qquad
\label{shh3}
\en
and 
\eq 
\quad
\Bigl[-\frac{d^2}{dx^2}+x^{6} + (2J {-} 2 l {-}1) x^{2}+ 
\frac{(J{+}l)(J{+}l{+}1)}{x^{2}} \Bigr]\Phi(x)=E\,\Phi(x) \, ,
\quad\Phi|_{x \rightarrow 0} \sim x^{J+l+1}\,.
\qquad
\label{shh4}
\en 
If $J$ is a negative integer, the mapping still makes sense, but it acts in
the opposite sense: shifting $l\to l{-}J$, (\ref{shh4}) becomes the QES
problem for $\alpha_{|J|}$, and (\ref{shh3}) its partner with the QES levels
removed.

Finally, we still have 
the freedom to apply the `tilde-duality' $\TT$
discussed in \S 5, so (\ref{shh4}) is in turn isospectral to
\eq 
\quad
\Bigl[-\frac{d^2}{dx^2}+x^{6} + (2J {+} 4 l{+}2) x^{2}+ 
\frac{(J{-}\frac{1}{2})(J{+}\frac{1}{2})}{ x^{2}} \Bigr]\Phi(x)=E\,\Phi(x) \, , 
\quad\Phi|_{x \rightarrow 0} \sim x^{J-1/2+1}\,.
\qquad
\label{shh5}
\en
This chain of equivalences will be discussed further in the conclusions.

Since the quasi-exactly solvable energies and the associated
wavefunctions are in principle exactly known,
one could eliminate them
one by one using the Darboux transformation, though this would be a
lengthy business for large values of $J$.
What seems surprising about the results (\ref{shh4}) and (\ref{shh5}) is 
that the  potential can have such a simple form once {\em all} of these levels
have been subtracted.
\resection{The fourth and fifth spectral equivalences}
The equivalence of the second-order equation with an $SU(3)$-related
third-order equation suggests two further spectral
equivalences. As explained in
\cite{DTc,DDT2}, the $\ZZ_2$ symmetry of the $SU(3)$ Dynkin diagram 
is reflected in a relation between the functions $y$ and $W$ that were
introduced in \S3. Explicitly,
\eq
y(x,E,\bg^{\dagger})
=W[y_{-1/2},y_{1/2}](x,E,\bg)\,,
\label{fff}
\en
where $\bg^{\dagger}=\{g_0^{\dagger},g_1^{\dagger},g_2^{\dagger})$ and
$g_i^{\dagger}=2{-}g\phup_{2-i}$\,.
On the second-order side of the story, a similar Wronskian appears, but this
time in the formula (\ref{tform}) for $T(E,\alpha,l)$:
\eq
T(E,\alpha,l)=W[\,Y_{-1}\,, Y_1](E,\alpha,l)\,.
\label{lll}
\en
Before the two equations can be compared, the $x$-dependence must be
eliminated from (\ref{fff}), and as usual this is done by considering the
behaviour as $x\to 0$. Extending the definition (\ref{YWnear0}), we define
three
functions $D^{(1)}_{[i]}$, with $D_{\phantom{[]}}^{(1)}\equiv D^{(1)}_{[0]}$, by
\eq
y(x,E,\bg)=\sum_{i=0}^2D^{(1)}_{[i]}(E,\bg)\,\chi_i(x,E,\bg)\,,
\en
where the solutions $\chi_i(x,E,\bg)$ to (\ref{invH})
are defined by $\chi_i\sim x^{g_i}+O(x^{g_i+3})$ as $x\to 0$.
To match (\ref{lll}), we expand out the RHS of (\ref{fff}) and then project
onto the component
behaving as $x^{g_0+g_2-1}=x^{g_1^{\dagger}}$ to find
\bea
D^{(1)}_{[1]}(E,\bg^{\dagger})\!&=&\!
(g_2{-}g_0) \Bigl[\, 
\omega^{(g_0-g_2)/2}
D^{(1)}_{[0]}(\omega^{3N/2}E,\bg) D^{(1)}_{[2]}(\omega^{-3N/2}E,\bg)
\qquad\qquad\quad\nn\\[3pt]
&&\qquad\qquad\quad {}-\,
\omega^{(g_2-g_0)/2}
D^{(1)}_{[0]}(\omega^{-3N/2}E,\bg) D^{(1)}_{[2]}(\omega^{3N/2}E,\bg)
\,\Bigr]\,.
\label{eqab}
\eea
On the other hand, the function $Y(x,E,\alpha,l)$ can be expanded as
\eq
Y(x,E,\alpha,l)=D(E,\alpha,l)X(x,E,\alpha,l)+
D(E,\alpha,-1{-}l)X(x,E,\alpha,-1{-}l)
\en
with $X(x,E,\alpha,l)\sim x^{-l}$ as $x\to 0$. (Cf.\ eq.~(5.2) of
\cite{DTb}, but note
that the definition of $D(E,l)$ used in \cite{DTb} differs from
that used here by a factor of $(2l{+}1)^{-1}$.\,) Now substitute into
(\ref{lll}) taken at $(-E,-\alpha,l)$\,:
\bea
T(-E,-\alpha,l)
\!&=&\!
(2l{+}1) \Bigl[\, 
\Omega^{-2l{-}1}
D(-\Omega^{2M}E,\alpha,l) D(-\Omega^{-2M}E,\alpha,-1{-}l) 
\qquad\qquad\qquad\nn\\[3pt]
&&\qquad\qquad\quad {}-\,
\Omega^{2l{+}1}
D(-\Omega^{-2M}E,\alpha,l) D(-\Omega^{2M}E,\alpha,-1{-}l) 
\,\Bigr]\,.
\label{eqac}
\eea
If $M=3$ and $N=1$, then $-\Omega^{2M}=\omega^{3N/2}=e^{\pi i/2}$, and,
if $\bg$ and $l$ are related by (\ref{alphag}),
$\omega^{(g_0-g_2)/2}=\Omega^{-2l{-}1}=e^{-\pi i(2l{+}1)/4}$.
Furthermore, from (\ref{prop}),
\eq
D^{(1)}_{[0]}(E,\bg)= f(\alpha,l)\, D(\kappa E,\alpha,l)~~,~~
D^{(1)}_{[2]}(E,\bg)= f(\alpha,-1{-}l)\, D(\kappa E,\alpha,-1{-}l)~.
\en
(The second relation is obtained by a continuation which swaps $g_0$ and
$g_2$.) Using these identifications and comparing (\ref{eqab}) and
(\ref{eqac}) gives our fourth spectral equivalence:
\eq
D^{(1)}_{[1]}(\kappa^{-1}E,\bg^{\dagger})
= \frac{3}{2} f(\alpha,l)f(\alpha,-1{-}l)\,T(-E,-\alpha,l)\,,
\label{spec4}
\en
where $g^{\dagger}_i=2-g^{\phantom{\dagger}}_{2-i}$ and
$\alpha=2(2-g_0-g_2)$, $l=(2g_2-3-2g_0)/6$\,. As with the first equivalence,
this relates spectral data for differential
equations of different orders. The spectral
interpretation of functions such as $T$ in the ODE/IM correspondence was 
discussed in \cite{DTb}, and is reviewed and extended to the current context
in appendix~B below. 

We can obtain a relation between objects in the second-order equation by
using the first spectral equivalence to rewrite the LHS of (\ref{spec4}), and
this constitutes our fifth and final spectral equivalence.
The only subtlety is that 
(\ref{prop}) involves $D^{(1)}_{[0]}$, not $D^{(1)}_{[1]}$, and this can be
overcome by a continuation in the $g_i$\,. Swapping $g_0$ and $g_1$ and
tracing back,
\eq
T(-E,-\alpha,l) \,=\, \frac{2f(-\wt\alpha,\wt l\,)}{3f(\alpha,l)f(\alpha,-1{-}l)}\,
D(E,-\wt\alpha(\alpha,l),\wt l(\alpha,l))\,.
\label{spec5}
\en
The proportionality factor can also be found explicitly, by considering
(\ref{spec5}) at $E=0$ and using formulae (\ref{Dzero}) and (\ref{Tzero}). The
result:
\eq
T(-E,-\alpha,l) \,=\, 
\frac{2\,\sqrt{i\pi}}
{\Gamma\bigl(\,\wt l(\alpha,l)+\frac{1}{2}\,\bigr)}\,
D(E,-\wt\alpha(\alpha,l),\wt l(\alpha,l))\,.
\label{speca}
\en
Via the second equivalence, this can be rewritten as
\eq
T(-E,\alpha,l)\,=\,
\frac{2\,\sqrt{i\pi}}
{\Gamma\bigl(-\wt l(\alpha,l)-\frac{1}{2}\,\bigr)}\,
D(E,\wt\alpha(\alpha,l),-1{-}\wt l(\alpha,l))\,.
\label{specb}
\en

As explained in appendix~B, $T$ is the spectral determinant for
a `lateral connection' problem, with the wavefunction lying on a contour in
the complex plane joining a pair of Stokes sectors at infinity. 
In contrast, $D$ is the
spectral determinant for a `radial connection' problem, with the wavefunction
living on a half-line. Similar equivalences, albeit for slightly different
potentials, have been found in \cite{BG}, and it would be interesting to see
whether similar methods could be applied in this case.

The mappings of parameters involved in these relations can be streamlined by
introducing two further matrices, $\AAa$ and $\LL$, again acting on the vectors
$\balpha=(\alpha,l,1)^T$:
\eq
\AAa= \left(\begin{array}{rrr} 
-1&0&0\\
0&1&0\\
0&0&1
\end{array}\right)~~;\quad
\LL= \left(\begin{array}{rrr} 
1&0&0\\
0&-1&-1\\
0&0&1
\end{array}\right)~,
\en
and setting 
\eq
\gamma(\balpha)\equiv \gamma((\alpha,l,1)^T)=2\sqrt{i\pi}/\Gamma(l{+}\fract{1}{2})~.
\en
The tilde-duality (\ref{spec2}) is then
\eq
\gamma(\balpha)D(E,\balpha)=\gamma(\TT\balpha)D(E,\TT\balpha)\,,
\en
while
(\ref{speca}) and (\ref{specb}) are, respectively,
\eq
T(-E,\AAa\balpha) \,=\,
\gamma(\AAa\TT\balpha)\,D(E,\AAa\TT\balpha)~~,\quad
T(-E,\balpha) \,=\,
\gamma(\LL\TT\balpha)\,D(E,\LL\TT\balpha)~.
\label{tgam}
\en
Note also that $\HH=\AAa\TT\AAa$, so the first
relation of (\ref{tgam})  can be rewritten as
$T(-E,\balpha)=\gamma(\HH\balpha)D(E,\HH\balpha)$. At the QES points,
$\balpha=\balpha_J=(\alpha_J(l),l,1)^T$, the spectrum encoded by
$D(E,\HH\balpha_J)$ is equal to that of $D(E,\balpha_J)$, apart from the QES
levels. As will be explained in the next section, these levels are in fact 
the zeroes of the Bender-Dunne polynomial $P_J(E)$, so at the QES points we have
\eq
P_J(E)T(-E,\balpha_J)\propto D(E,\balpha_J)~,
\label{ptd}
\en
which is a relation between spectral data for lateral and
radial connection problems with the {\em same} Hamiltonian.

The algebra of the matrices we have introduced is best described by
first defining $\MM=\AAa\LL$. Then a set of defining relations for $\LL$, $\MM$
and $\TT$ is
\eq
\LL^2=\MM^2=\TT^2=(\LL\MM)^2=(\MM\TT)^2=(\LL\TT)^3={\mathbb I}~.
\en
Thus $\MM$ commutes with $\LL$ and $\TT$, while $<\LL,\TT>$ forms the Weyl
group of $SU(3)$. However, in general only $\TT$ yields a spectral equivalence
of the $D(E,\balpha)$. We will return to this point in the conclusions.

We end this section with two further remarks about the fifth set of
equivalences. First, they  can
also be obtained entirely in the context of the second-order differential
equation. For $M=3$, 
manipulating equation (\ref{tqa}) and using the fact that $\Omega^{2M}=-i$
leads to the following  functional relation, special to this particular value
of $M$:
\bea
\hspace{-1.cm}
2 \sin(\fract{\pi}{4}(2l{+}1{-}\alpha)) D^{(+)}(E) &=& \nn \\
&& \hspace{-4.cm} 
\Omega^{(2l+1-\alpha)/2}  
T^{(-)}(-iE) D^{(-)}(-iE)   
-  \Omega^{-(2l+1-\alpha)/2}  T^{(-)}(iE)D^{(-)}(iE)  
 \, .
\label{TDTD}
\eea
{}Taking (\ref{TDTD}) at $E=E^{(+)}_k$\,, combining it with (\ref{tqa}),
also at  $E=E^{(+)}_k$, and finally expressing the result in a factorised form
over the zeroes of $T^{(-)}(E)$  (which we denote by 
$\{ -\lambda^{(-)}_k\}$\,)  yields the following set of constraints:
\eq
\prod_{n=0}^{\infty} \lf( { \lambda^{(-)}_n - \Omega^{-2M} \, 
E^{(+)}_k   \over 
\lambda^{(-)}_n - \Omega^{\;2M} \, E^{(+)}_k }
\ri)
= - \Omega^{-4l-2}\ . \label{batq0}  
\en
A complementary set arises from (\ref{tq}), taken at $E=-\lambda^{(-)}_k$\,:
\eq
\prod_{n=0}^{\infty} \lf( { E^{(+)}_n - \Omega^{-2M} \, \lambda^{(-)}_k  \over 
E^{(+)}_n -  \Omega^{\;2M} \, \lambda^{(-)}_k}
\ri)
= - \Omega^{2l+1 - \alpha } \, . \label{batq1} 
\en
Together, (\ref{batq0}) and (\ref{batq1}) form a set of Bethe ansatz equations
of exactly the same form as
(\ref{ba01}) and (\ref{ba02}), save for the replacement of $\alpha$
and $l$ on the right-hand sides of (\ref{ba01}) and (\ref{ba02}) by
$\wt{\alpha}(\alpha,l)$ and  $\wt{l}(\alpha,l)$, respectively. 
By comparing the left-hand sides and 
exploiting the analytic
properties derived in  appendix~B, one  obtains by another
route the
second and fifth equivalences (\ref{spec2}), (\ref{spec5}):
\eq
E^{(+)}_k(\alpha,l) = E^{(+)}_k(\wt{\alpha},\wt{l})~~~\mbox{and}~~~ 
\lambda^{(-)}_k(\alpha,l) =  E^{(-)}_k(\wt{\alpha},\wt{l}) \, .
\en
The previous approach, which proceeded via the symmetries of the third-order 
equation, was perhaps more elegant. The advantage of this alternative method is
that the only  analytic properties used are those of spectral
determinants of the second-order equation, and these are known rigorously
from the results in appendix~B.

The second remark relates to the fact that the
lateral connection problem solved by $T$ is closely related to ${\cal
PT}$-symmetric quantum mechanics. These problems are not, in any obvious
sense, self-adjoint, and the reality properties of their
spectra have long been of interest 
\cite{BZJ,BB,BBM,DP,DT,Mez,Zl,BCQ,BDMS,BW,Mez1}. Since 
the spectral problems on the right-hand sides of (\ref{speca}) or
(\ref{specb}) {\em are} self-adjoint for $\wt l>-1/2$ (respectively 
$-1{-}\wt l>-1/2$), these two identities give us a simple way to understand the 
reality of the spectra encoded by $T$ in these particular cases. However, 
in appendix B below we will give a proof of reality which is both
more general and more direct, and so we will not 
pursue this any further.

\resection{A relation with Bender-Dunne polynomials}
The dualities that we have been discussing have an interesting relationship
with the so-called Bender-Dunne polynomials. These were introduced in \cite{BD}
as a way of understanding quasi-exact solvability, but here we will also be
interested in their properties at general values of the parameters. Briefly, one
searches for a solution to (\ref{x6}) of the form
\eq
\psi(x,E,\alpha,l)=
e^{-x^4/4}\,
x^{l{+}1}\,
\sum_{n=0}^{\infty}
\left({\scriptstyle -}\fract{1}{4}\right)^{n}
\frac{P_n(E,\alpha,l)}{n!\,\Gamma(n{+}l{+}3/2)}\,x^{2n}\,.
\label{bdser}
\en
For this to solve the differential equation
(\ref{x6}), the coefficients $P_n$ must satisfy the following 
recursion relation:
\eq
P_n(E)= E P_{n-1}(E) + 16(n-1)(n-J-1)(n+l-1/2)P_{n-2}(E)\, ,~~~(n \ge 1)
\label{rec}
\en
where, as before, $J=J(\alpha,l)=-(\alpha{+}2l{+}1)/4$.
The value of $P_0(E)$, which determines  
the normalisation of $\psi(x)$, is conventionally taken to be $1$; from (\ref{rec}),
$P_1=E$, and $P_n$ is a polynomial of degree $n$ in $E$, known as a Bender-Dunne 
polynomial. So long as $l\neq -n{-}3/2$ for any $n\in\ZZ^+$,
(\ref{bdser}) will yield
an everywhere-convergent series solution to (\ref{x6}).
Furthermore, this solution automatically satisfies the boundary condition
$\psi \sim x^{l+1}$ at $x = 0$\,; but at general values of $E$, it will grow
exponentially as $x\to\infty$.
We now ask whether there are transformations of the parameters $\alpha$
and $l$
which leave the Bender-Dunne polynomials invariant. It is easily seen 
that if $J$ and $l$ are replaced by $\wt J=-l-1/2$ 
and $\wt l=-J-1/2$, then the recursion relation is
unchanged. Translated back to the parameters $\alpha$ and $l$,
this implies that
\eq
P_n(E,\alpha,l)=P_n(E,\wt\alpha,\wt l)
\label{pres}
\en
where
\eq
\wt\alpha=2J+4l+2=3/2-\alpha/2+3l~,\quad
\wt l =-J-1/2= \alpha/4+l/2-1/4~.
\en
This matches the `second spectral equivalence' found earlier. 

We will return to this
case later, but first we discuss
the special points where the model is quasi-exactly solvable, for which
a similar game
can be played. 
If $\alpha$ and $l$ are such that $J(\alpha,l)$ is a positive integer,
the second term on the RHS of
(\ref{rec}) vanishes at $n=J{+}1$,
and
all the subsequent $P_n$ therefore factorise:
\eq
P_{n{+}J}(E,\alpha_J,l)=P_J(E,\alpha_J,l)Q_n(E,\alpha_J,l)~,
\quad (n>0, J=-(\alpha_J{+}2l{+}1)/4\in \NN).
\en
Hence, if $P_J(E)$ vanishes then so do all $P_{n\ge J}(E)$ and the series
(\ref{bdser}) terminates, automatically yielding a normalisable solution to
(\ref{x6}). The $J$ zeroes of $P_J(E)$ are the $J$ exactly-solvable levels for
the model and, as observed by Bender and Dunne, this provides a simple way to
understand the quasi-exact solvability of the model. Now we would like to go further
and discuss the remaining levels. The polynomials $Q_n$ satisfy the recursion
\eq
Q_n(E)= E Q_{n-1}(E) + 16(n+J-1)(n-1)(n+J+l-1/2)Q_{n-2}(E)\, ,~~~(n \ge 1)
\label{qrec}
\en
with initial conditions $Q_0=1$, $Q_1=E$. This matches the recursion
relation for $P_n(E)$, so long as $J$ and $l$ in (\ref{rec}) are replaced
by $\widehat J=-J$ and $\widehat l=J+l$.
Hence, if
\eq
\widehat\alpha_J=2J-2l-1=-\alpha_J/2-3l-3/2~,\quad
\widehat l_J = J+l=-\alpha_J/4+ l/2-1/4~,
\en
then
\eq
Q_n(E,\alpha_J,l) =P_n(E,\widehat\alpha_J,\widehat l_J)~.
\label{qp}
\en
This corresponds to the `third spectral equivalence', and it
has an interesting consequence for the series expansion (\ref{bdser}),
which we rewrite as
\bea
\psi(x,E,\alpha_J,l)&=&
e^{-x^4/4}\,
x^{l{+}1}\, \left[~\dots\,+
\sum_{n=J}^{\infty} \left({\textstyle -}\fract{1}{4}\right)^{n}
\frac{P_n(E,\alpha_J,l)}{n!\,\Gamma(n{+}l{+}3/2)}\,x^{2n}\,\right]\nn\\[3pt]
=~&&
\!\!\!\!
\!\!\!\!
\!\!\!\!
e^{-x^4/4}\, x^{l{+}1}\, \left[~\dots\,+
\sum_{n=0}^{\infty} \left({\textstyle -}\fract{1}{4}\right)^{n+J}
\frac{P_J(E,\alpha_J,l)Q_n(E,\alpha_J,l)}{(n{+}J)!\,
\Gamma(n{+}J{+}l{+}3/2)}\,x^{2(n{+}J)}\,\right]
\qquad
\quad
\label{bdrser}
\eea
the dots standing for lower-order terms. 
This can be compared with the expansion of the wavefunction
$\psi(x,E,\widehat\alpha_J,\widehat l_J)$. Using $\widehat l_J=J{+}l$ and the
equality (\ref{qp}), this is
\eq
\psi(x,E,\widehat\alpha_J,\widehat l_J)=
e^{-x^4/4}\,
x^{l{+}J{+}1}\,
\sum_{n=0}^{\infty}
\left({\scriptstyle -}\fract{1}{4}\right)^{n}
\frac{Q_n(E,\alpha_J,l)}{n!\,\Gamma(n{+}J{+}l{+}3/2)}\,x^{2n}\,.
\label{bdserhat}
\en
It is now easy to see that $\psi(x,E,\alpha_J,l)$ is mapped onto a function
proportional to $\psi(x,E,\widehat\alpha_J,\widehat l_J)$
by the differential operator
\eq
\CQJ(l)\, =\, e^{-x^4/4}x^{l+J+1}\left(\frac{1}{x}\frac{d}{dx}\right)^J
e^{x^4/4}x^{-l-1}
\,=\,x^J\left[
\frac{1}{x}\frac{d}{dx} +x^2-\frac{l{+}1}{x^2}
\right]^J \, .
\en
This is enough to see that the following intertwining relation between 
differential operators must hold:
\eq
\CQJ(l)\,\CH(\alpha_J(l),l)=\CH(\widehat\alpha_J(l),\widehat
l_J(l))\,\CQJ(l) \, .
\label{intertw}
\en
(Consider\footnote{We would like to
thank Peter Bowcock for a discussion of this point.}
the difference between the LHS and RHS. This is a
linear $(J{+}2)^{\rm th}$-order
differential operator, independent of $E$, and it is easily seen that it
annihilates the functions $\psi(x,E,\alpha_J(l),l)$. These
functions are linearly independent for different
values of $E$, while a $(J{+}2)^{\rm th}$-order operator can annihilate at
most $(J{+}2)$ independent functions, unless it is identically zero. This
establishes the equality.) We also
used Maple to verify (\ref{intertw}) directly.
Finally, $\CQJ(l)$ respects the boundary conditions:
if $\psi(x)$ decays as $x\to\infty$ then so does $\CQJ(l)\,\psi(x)$\,, 
and if $\psi(x)$, given as a series by (\ref{bdser}), has leading behaviour
$x^{l+1}$ at the origin, then $\CQJ(l)\,\psi(x)$ has leading behaviour $x^{l+J+1}$.
Thus $\CQJ(l)$ maps eigenfunctions of the problem
$\CH(\alpha_J,l)$ to those of
$\CH(\widehat\alpha_J(l),\widehat l_J(l))$, or to zero. The
eigenfunctions mapped to zero are those for which $P_J(E,\alpha_J,l)$
vanishes (the lower-order terms `\dots' in (\ref{bdser})
are clearly annihilated by $\CQJ(l)$),
and these are precisely the exactly-solvable levels. This provides an alternative
derivation of the duality found with the aid of
 the Bethe ansatz equations in \S 6, and shows
that in $\CQJ$ we have found the generalisation of the
supersymmetry operator $\CQ^+\equiv \CQ\phup_1$ to the QES problems 
(\ref{shh3})
with $J>1$.

So far we have discussed the action of $\CQJ(l)$ on solutions to the spectral
problem with $x^{l+1}$ boundary conditions. But given the intertwining relation
(\ref{intertw}) it is natural to look for an action on solutions 
satisfying the
other, $x^{-l}$, boundary condition at the origin. 
As a relation between differential operators, the fact that
$\CH(a,b)=\CH(a,-1{-}b)$ means, trivially, that 
\eq
\CQJ(l)\,\CH(\alpha_J(l),{-}1{-}l)=
\CH(\widehat\alpha_J(l),{-}1{-}\widehat l_J(l))\,\CQJ(l) \, .
\label{intertwa}
\en
It can also be checked that, in general, the relevant boundary conditions are
respected, so that (\ref{intertwa}) holds as an intertwining relation
between eigenvalue problems.
Substituting $-1{-}l$ for $l$ throughout,
the relation is
\eq
\CQJ(-1{-}l)\,\CH(\alpha_J(-1{-}l),l)=
\CH(\widehat\alpha_J({-}1{-}l),{-}1{-}\widehat l_J({-}1{-}l))\,\CQJ({-}1{-}l)\,.
\label{intertwd}
\en
Thus $\CQJ(-1{-}l)$ and its adjoint intertwine 
between the spectral problems
\eq
\CH(-4J{+}2l{+}1,l)
\mbox{~~and~~}
\CH(2J{+}2l{+}1,l{-}J)\,,
\label{intertwe}
\en
and, in general, 
no eigenfunctions are annihilated. Furthermore,  the mapping 
(\ref{intertwe})
is exactly the `second spectral equivalence' of \S5 above,
specialised to cases where the initial pair of parameters $(\alpha,l)$ satisfies
$\alpha=-4J+2l+1$. This hints at an alternative way to obtain
(\ref{intertwd}): just as was done at the QES points using (\ref{qp}), one
can compare the series expansions 
using (\ref{pres}). At a formal level,
for any value of $\alpha$ and $l$ the series for $\psi(x,E,\alpha,l)$ is
mapped onto that for $\psi(x,E,\wt\alpha,\wt l)$ by 
\eq
{\cal P}_{\mu}(l) \, =\, 
e^{-x^4/4}\,x^{\mu-l}
\left(\frac{1}{x}\frac{d}{dx}\right)^{\mu}\,
e^{x^4/4}x^{l}\, ,
\en
where $\mu = -\alpha/4+l/2+1/4$, and the action of a fractional power of
the derivative is, again formally, defined by
\eq
\left(\frac{1}{x}\frac{d}{dx}\right)^{\mu}x^{2n}=
2_{\phantom{l}}^{\mu}\frac{\Gamma(n{+}1)}{\Gamma(n{+}1{-}\mu)}\,x^{2(n{-}\mu)}.
\en
In cases when $\mu$ is a positive integer, ${\cal P}_\mu(l)$ becomes an
ordinary differential operator, equal to $\CQ_{\mu}({-}1{-}l)$, and 
(\ref{intertwd}) is recovered. It is interesting to speculate about the
existence of some kind of spontaneously-broken
`fractional' generalised supersymmetry lying
behind the tilde-duality at arbitrary values of the
parameters, but we leave this for future
work.

In concluding this section, we would like to mention the recent article
\cite{nfold}, which we noticed as we were finishing the writing of this
paper. By a completely different route, involving a study of a concept
called `${\cal N}$-fold supersymmetry' \cite{AIS,nfoldo}\footnote{generalised 
supersymmetries of this sort are
also called `higher-derivative', or `non-linear'~\cite{KP}.}
the authors of this work have
also introduced higher-order analogues of the supersymmetry generators.
Although the connection with quasi-exact solvability
is not mentioned,
one can check that the
`cubic' case of the type A ${\cal N}$-fold supersymmetry
 of \cite{nfold}
 reproduces the result
(\ref{intertw}) above, albeit with a slightly different presentation of
the operators. 
(In fact, using the most general form of their operators, one can also
obtain an intertwining relation for the more general QES sextic potentials
involving an additional $x^4$ term.) We should also mention that a connection
between certain other forms of
non-linear supersymmetry and quasi-exact solvability has recently been
pointed out in \cite{KP}. However the forms of the supersymmetry generators
explicitly treated in that paper do not cover the case of the sextic
potential discussed above.

\resection{Conclusions}
The main  purpose of this paper has been to illustrate how spectral properties 
and symmetries of  interesting differential operators can be 
handled using tools originally developed in the context of integrable
models. These lead to some novel spectral equivalences, 
and, as will be shown in appendix~B below, they also allow for an
elementary proof of a reality
property which has been surprisingly elusive when studied by
more conventional methods.
Some of the equivalences
we have subsequently been able to re-derive by other means,
and in this respect, the
r\^ole of higher-order generalisations of the supersymmetry operators 
is particularly intriguing, especially in the light of their independent
appearance in \cite{nfold}. It would be very interesting to find out whether
the connection between such operators and quasi-exact solvability that we
have observed is more general. In this paper, we obtained the operators
$\CQJ$ through a direct examination of power series solutions; how they 
fit into the more algebraic schemes for understanding quasi-exact
solvability, as developed in, for example, \cite{Tur,Ush}, is another 
question that deserves further study.

In many ways, the first and fourth spectral equivalences, 
between second- and third-order
equations, are the most unexpected of our results.
They can be traced back to the collapse of the $SU(3)$ Bethe ansatz equations
at $N{=}1$. A similar phenomenon
occurs in $SU(n)$-related
BA systems for $n>3$, which
are related to higher-order
differential equations 
via the ODE/IM correspondence~\cite{DTc,Sb,DDT2}.
However, in these cases the resulting
`reduced' systems are not so readily identifiable, and so at this stage we
lack an interpretation of the phenomenon in terms of the properties of
differential equations. On the other hand,
via the first equivalence we do at least see that
quasi-exact solvability is not restricted to second-order spectral
problems. 

One can also ask whether
quasi-exact solvability might have a r\^ole
on the `integrable models' side of the
ODE/IM correspondence.  Bethe roots correspond to zeroes of $D(E)$, and
at the QES points the locations of a finite subset of these can be found 
exactly. From the identity
(\ref{ptd}), the locations of the remaining roots coincide with the zeroes of
$T(-E)$, while, as
follows from the form of the
Bender-Dunne wavefunctions, the QES roots themselves correspond to
coincidences in the locations of
zeroes of $D^{(\pm)}(E)$ and $T^{(\mp)}(E)$. However, we do not know of any
special significance of these facts.

Finally, it is interesting to draw the full set of spectral problems
that can be reached from a QES starting-point using the dualities that we
have been discussing:
\bea
\CH(-4J{-}2l{-}1,l)~~~~~~ &\longrightarrow& ~~\,~\CH(2J{-}2l{-}1,J{+}l) \nn\\[7pt]
\Big\updownarrow \qquad\qquad~~~~&&
\qquad~~\qquad\Big\updownarrow \nn\\[9pt]
\CH(2J{+}4l{+}2,-J{-}\fract{1}{2})~~~ &\longrightarrow&~~~
\CH(2J{+}4l{+}2,J{-}\fract{1}{2}) \nn
\eea

Vertical arrows correspond to the second spectral equivalence $\TT$, while the
upper
horizontal arrow is the level-eliminating third equivalence, $\HH$. 
The two problems on the bottom row are related by the transformation
$\LL$: $l\to -1{-}l$. They correspond to the {\em same}
Schr\"odinger equation, and differ only in the boundary
condition imposed at the origin. 
It follows from the diagram that the `regular' eigenvalue problem
for this equation, that with the
$x^{J-1/2}$ boundary condition, has exactly the same spectrum as the
irregular problem, with the exception of the first $J$ levels.
This can be understood by noticing that, at $l=J{-}1/2$,
there is a `resonance' between the regular and
irregular solutions of the Schr\"odinger equation (see, for example,
\cite{NEWT}). The Bender-Dunne series expansions for the
first $J$ levels of the irregular problem truncate
before the resonance is reached, whilst the wavefunctions for the remaining 
levels are completely dominated by the effect of the resonance, and hence
match those of the regular problem. 
In fact, such a square of spectral problems can be drawn starting from 
{\em any} values of the parameters $\alpha$ and $l$, on account of the
identity $\HH=\TT\LL\TT$. But it is only at the QES points that
the horizontal directions correspond to (partial) spectral equivalences, 
since the resonance between regular and
irregular solutions just described only occurs when $J$ is an integer.
Thus we have some novel points at which the
sextic potential can be considered to be quasi-exactly solvable, a
`dual' interpretation of quasi-exact solubility for this model in terms of 
the resonance of irregular solutions of the Schr\"odinger equation,
 and an alternative interpretation
of the level-elimination at work in the `supersymmetric' third spectral
equivalence.
\medskip

%

\noindent{\bf Acknowledgements ---~} 
We would like to thank Sharry Borgan, Peter Bowcock, Francesco Cannata,
Davide Fioravanti, Brent Everitt, 
Giuseppe Mussardo, John Parker, Junji Suzuki, Alexander Turbiner,
Andr\'e Voros and Jean Zinn-Justin for useful conversations and help. 
RT also
thanks Ferdinando Gliozzi and the Department of Theoretical Physics 
of  Torino University
for kind hospitality during  the final stages of this project, and we all
thank Hideaki Aoyama, Carl Bender, Andrei Mezincescu and Miloslav Znojil for
helpful correspondence.
The work was supported in part by a TMR grant of the
European Commission, reference ERBFMRXCT960012.
RT was also supported by an EPSRC VF grant, number GR/N27330.
\goodbreak
%
%
\appendix

\resection{Solving radial Schr\"odinger equations using Maple}
In this appendix we discuss the numerical treatment of the
radial Schr\"odinger equation. The standard method is to integrate the 
ODE at varying values of energy, imposing boundary conditions 
either at the origin or infinity, and
searching for the values of energy at which the
other boundary condition is satisfied (see for example \cite{LS}).
This approach runs into problems
in the region $\Re e\, l<-1/2$, where the eigenvalue problem is better defined
via analytic continuation. 
The series solution alternative that we describe here avoids this
difficulty, can be implemented in only a few lines of Maple and seems to be a rather
efficient method for finding the lower-lying levels, at least for a polynomial
potential.
We chose not to use a series of the sort described in \S 8 above,
as the factor of $\exp(-x^4/4)$ means that any finite
truncation of the series always decays at $x\to\infty$; it is only in the
infinite sum that the exponential growth of a solution at generic $E$ is 
recovered. This makes it hard to detect eigenvalues reliably.
Instead, we generated a pure
power series directly in Maple, using an algorithm based on
the method of Cheng \cite{cheng}. The power series {\tt y}
produced by the program depends both on {\tt x} 
and on {\tt E}, and by construction it satisfies the boundary condition at
{\tt x=0}. At an eigenvalue, the solution must decay at large {\tt x}, and by 
choosing a suitably-large value {\tt x}$_0$ and searching for 
values of {\tt E} at which
{\tt y}(x$_0${\tt ,E)=0}, the eigenvalues can be located with high accuracy.
The value of {\tt x}$_0$ must be large enough that the asymptotic behaviour of the
true solution has set in, but small enough that the approximated power series 
can be relied on.
(The level of the approximation is controlled by the variable
{\tt iterations} in the program below.) 
This can be checked by examining plots of
the candidate wavefunctions.
We first give the code that we used,
the particular example producing figure \ref{pa1} below.
The values of $\alpha$ and $l$ are specified in
the second line.\\[-17pt]

{\tt\small
\begin{verbatim}
> Digits:=20: iterations:=40:
> alpha:=-4: l:=0:
> V:=x^6+alpha*x^2;
> L:=poly->sum(coeff(poly,x,n)*x^(n+2)/(n+2)/(n+2*l+3),n=0..degree(poly,x)):
> P:=1:for i from 1 to iterations do P:= simplify(1+L((V-E)*P)) end do:
> y:=x^(l+1)*P:
> spectrum:=fsolve(eval(y,x=3.2)=0);
\end{verbatim}
}

\noindent
A specific level, determined
by the integer {\tt plotlevel},
is then plotted as follows:\\[-17pt]

{\tt\small
\begin{verbatim}
> with(plots): plotlevel:=0:Ee:=spectrum[plotlevel+1];
> xmax:=2.5:ymin:=-35:ymax:=100:
> display([
  plot(eval(80*y,E=Ee),x=0..xmax,ymin..ymax,color=blue,linestyle=2,thickness=2),
  plot((V+l*(l+1)/x^2),x=0..xmax,ymin..ymax,color=red,thickness=2),
  seq(plot([[0.05,spectrum[lev]],[0.12,spectrum[lev]]],
             color=black,linestyle=1,thickness=3), lev=1..7),
  plot([[0.01,spectrum[plotlevel+1]],[0.17,spectrum[plotlevel+1]]],
        color=black,linestyle=1,thickness=1)
  ]);
\end{verbatim}
}

\noindent
The levels are contained in the list {\tt spectrum}, and in table \ref{tabdata}
below we compare semiclassical, nonlinear integral equation (NLIE) and
Maple results for two sides of 
the $SU(3)$-inspired duality of \S5. To compile the table, we increased 
{\tt Digits} to $40$ and {\tt iterations} to $50$; nevertheless, each column
of data still took less than 6 minutes of CPU time on a 650 MHz Pentium III
machine, running under Linux.


\begin{table}[ht]
 \renewcommand{\arraystretch}{1.3}
\hskip -25pt
\begin{tabular}{ |r || c| c || c | c|  }                          \hline 
   & \multicolumn{2}{|c||} {$(\alpha=0.6,l=0.3)$} & \multicolumn{2}{|c|}
{$(\alpha=2.1,l=0.05)$}  \\  \hline
 \multicolumn{1}{|c||}{$E_{\rm sc}$}  &    $E_{\rm NLIE}$ & $E_{\rm
Maple}$   &   $E_{\rm NLIE}$  &  $E_{\rm Maple}$  \\
\hline
    5.9321  &    5.968835071930586       &            5.968835071930586
&   5.968835071930586   &  5.968835071930586   \\
    17.4454    &          17.45222031394444       &   17.45222031394444
&   17.45222031394444  &   17.45222031394444  \\
    32.4611    &        32.46443928842889         &   32.46443928842889
&   32.46443928842892  &   32.46443928842889   \\
    50.2951    &            50.29711627959066     &   50.29711627959066
&    50.29711627959066  &   50.29711627959066  \\
    70.5566    &             70.55798308714998    &   70.55798308714998
&   70.55798308714998  &   70.55798308714998   \\
    92.9834    &           92.98438725677185      &   92.98438725677185
&   92.98438725677185  &   92.98438725677184  \\
   117.3836   &    117.3842907593505      &           117.3842907593502
&   117.3842907593505  &   117.3842907593506   \\
   143.6087   &    143.6092804003097      &           143.6092804003098
&   143.6092804003097  &    143.6092804003097 \\
   171.5397   &    171.5401630800539      &           171.5401630800552
&   171.5401630800539  &    171.5401630800544  \\
   201.0781   &    201.0784630892047       &           201.0784630892253
&    201.0784630892047  &   201.0784630892116   \\
   232.1408  &     232.1410302822398  &           232.1410302455113
&   232.1410302822398&  232.1410302776264\\
   264.6562  &     264.6564313215198      &           264.6565101370171
&    264.6564313215198  &    264.6563953856096\\
   298.5623 &    298.5624300607573   &                299.0784472732795 &
298.5624300607573     &  298.7522851488036\\
   333.8040  &     333.8041674124193  &               339.1746343166020 &
333.8041674124193     &  347.4041682854545 \\
  \hline                                     
  \end{tabular}                                 
   \caption{ Numerical Results }      
  \label{tabdata}                                  
\end{table} 


{}From the table, it is clear that
the NLIE (\ref{nlies}) is able to find the 
energy levels with high accuracy, provided  $|\alpha| <M+1-|2l+1|$, and 
for such values of the parameters it seems to be the most reliable method
to find the full set of energy levels.
The power series approach also works extremely well for
the low-lying energy levels, but loses accuracy for the higher levels,
at least if the
value of {\tt iterations} is kept reasonably small.
Nevertheless, combining power series 
and semiclassical methods allows us to obtain good results, over the full
spectrum, for any
values of $\alpha$ and $l$.

\medskip

The figures which follow illustrate some typical wavefunctions.
In each case the low-lying energy levels are shown as bars
near the left-hand side 
of the plot, with the level under scrutiny having double length. The dashed
line is the (un-normalised)
wavefunction associated with this energy level, and the solid line
the potential itself.

The first set of four figures illustrates the second spectral
equivalence, connecting the 
potentials with $\alpha,l = -4,0$ and $7/2,-5/4$ respectively. Note that the
dual problem is an example where the eigenfunction is not
square-integrable; nevertheless, 
the numerical solution converges and reproduces exactly the required
spectrum.  Figures \ref{pa1} and \ref{pa2} show the 
ground state, and figures \ref{ppa1} and \ref{ppa2} 
the first excited state.

\bigskip

\[
\begin{array}{ll}
\refstepcounter{figure}
\label{pa1}
\includegraphics[width=0.460\linewidth,height=0.46\linewidth]{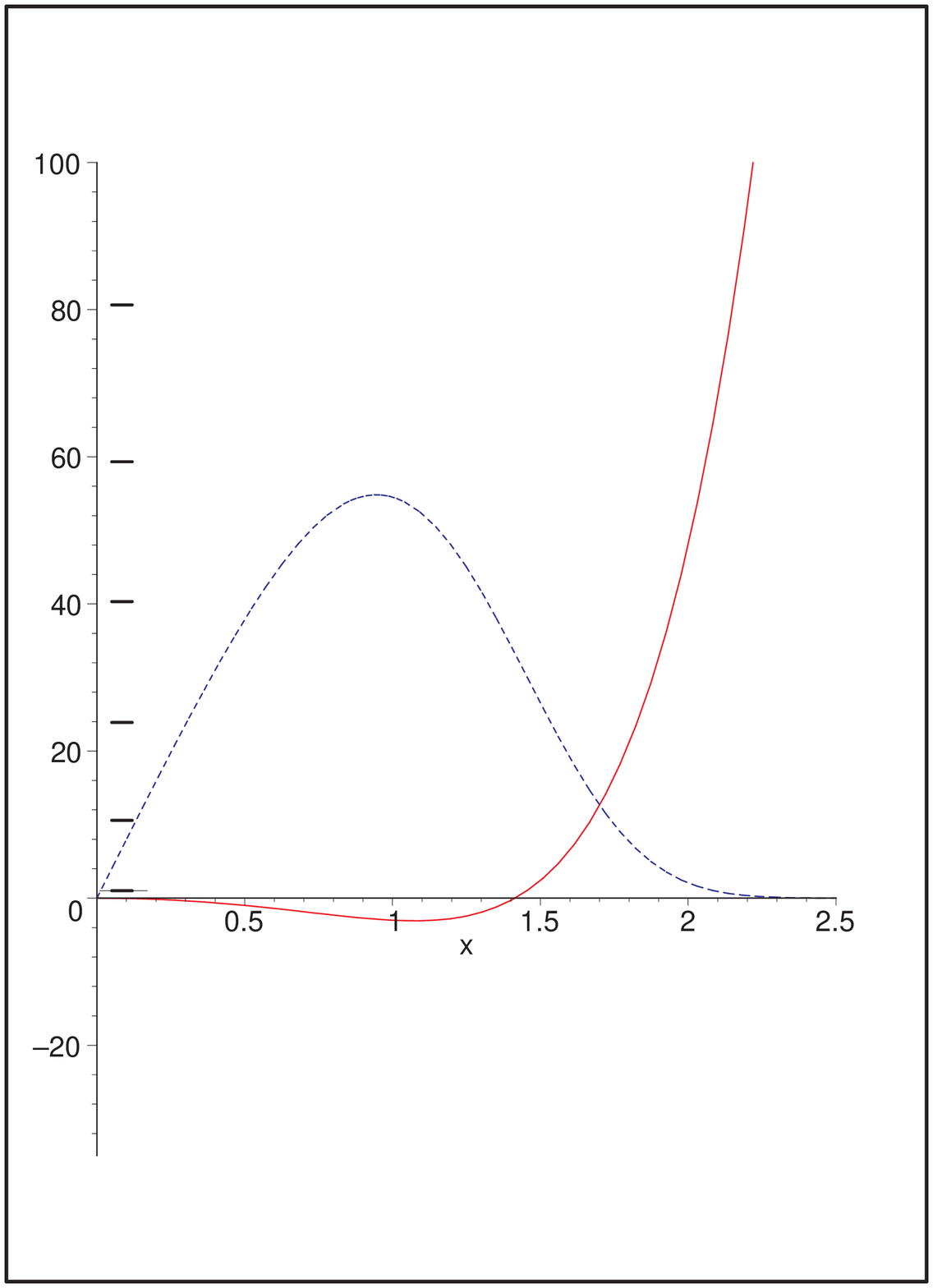}~~~~
&
\refstepcounter{figure}
\label{pa2}
\includegraphics[width=0.460\linewidth,height=0.46\linewidth]{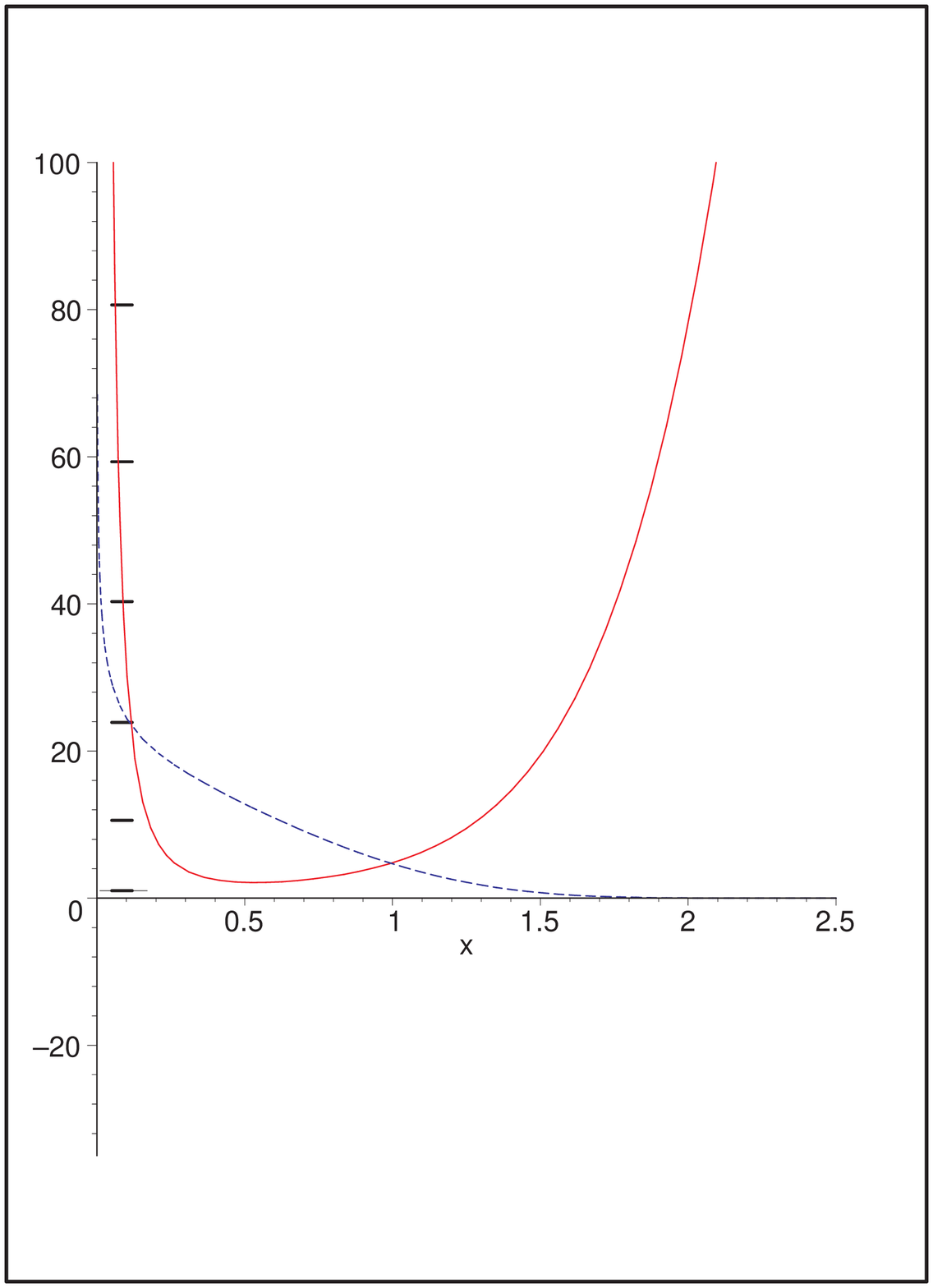}
\\
\parbox{0.45\linewidth}{
\raggedright
{\small Figure \ref{pa1}:
$\alpha=-4$, $l=0$, $E_0=1.0057683$
}}
&
\parbox{0.5\linewidth}{
\raggedright
{\small Figure \ref{pa2}:
$\alpha=7/2$, $l=-5/4$, $E_0=1.0057683$
}}
\end{array}
\]
\[
\begin{array}{ll}
\refstepcounter{figure}
\label{ppa1}
\includegraphics[width=0.460\linewidth,height=0.460\linewidth]{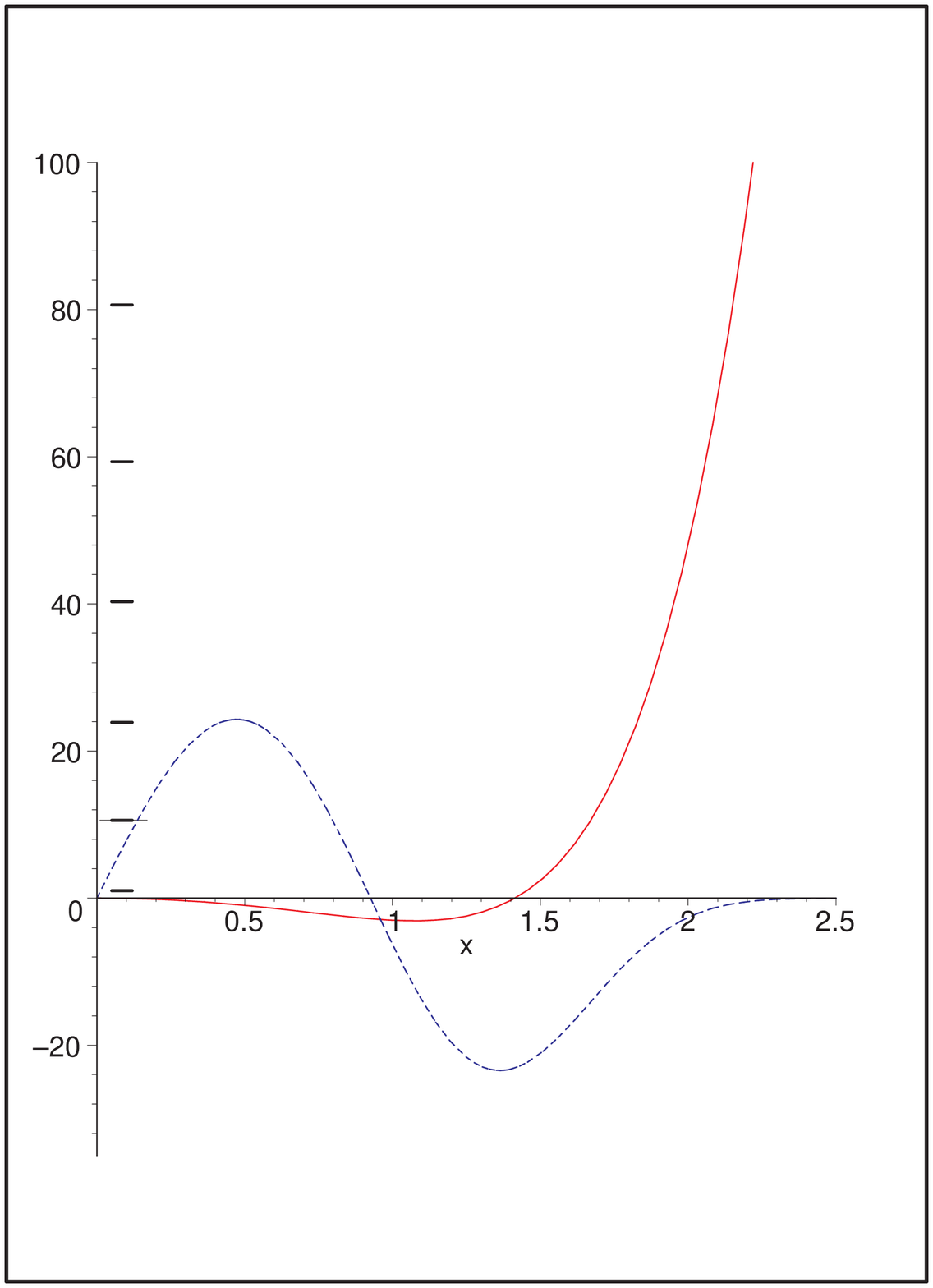}~~~~
&
\refstepcounter{figure}
\label{ppa2}
\includegraphics[width=0.460\linewidth,height=0.460\linewidth]{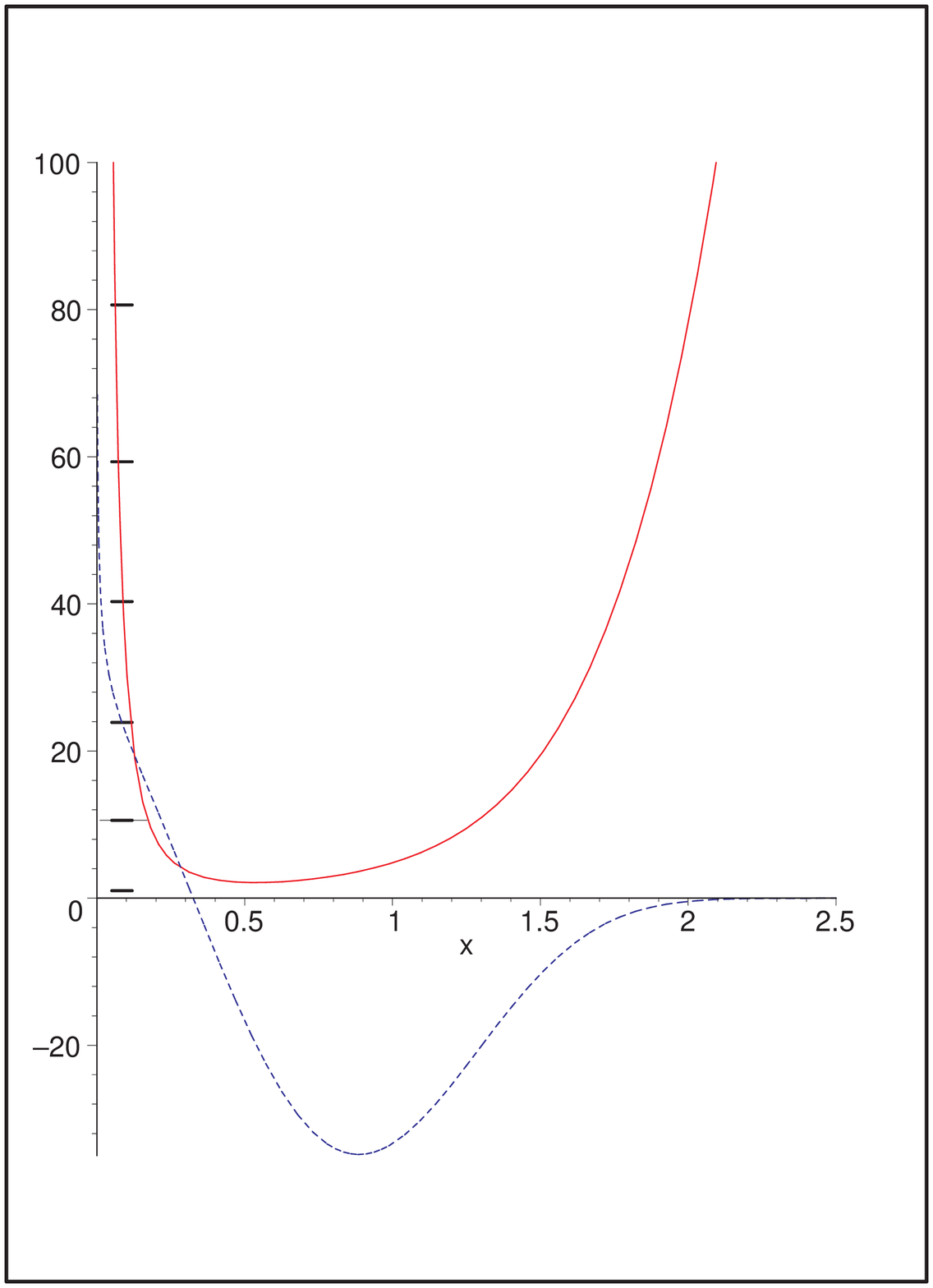}
\\
\parbox{0.45\linewidth}{
\raggedright
{\small Figure \ref{ppa1}:
$\alpha=-4$, $l=0$, $E_1=10.572585$
}}
&
\parbox{0.5\linewidth}{
\raggedright
{\small Figure \ref{ppa2}:
$\alpha=7/2$, $l=-5/4$, $E_1=10.572585$
}}
\end{array}
\]

\newpage

Figure \ref{p1}
depicts the quasi-exactly solvable case for the third energy level, at
$J=2$ and $l=0$. This corresponds to the ground state of 
the SUSY ($\HH$)
dual potential, shown in figure \ref{p2}. The fourth level of the QES
problem has the same energy as the first excited state of the SUSY 
dual, and
these two are shown in figures \ref{pp1} and \ref{pp2}. Figures
\ref{p3} and~\ref{pp3} then show the first two states of the potential related to
the SUSY dual by the second spectral equivalence, the tilde-duality $\TT$.

\medskip

%
\[
\begin{array}{ll}
\refstepcounter{figure}
\label{p1}
\includegraphics[width=0.460\linewidth,height=0.460\linewidth]{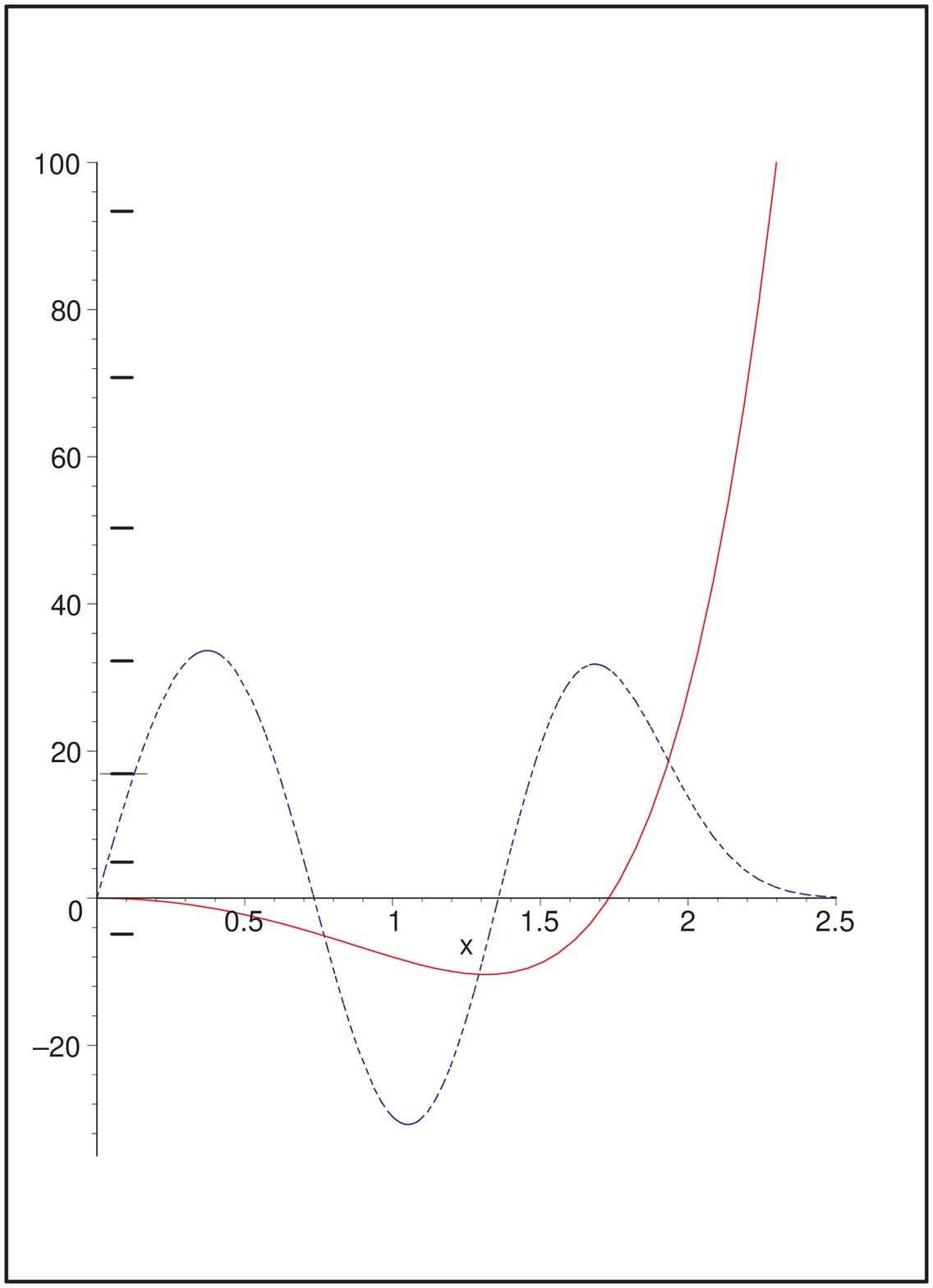}~~~~
&
\refstepcounter{figure}
\label{p2}
\includegraphics[width=0.460\linewidth,height=0.460\linewidth]{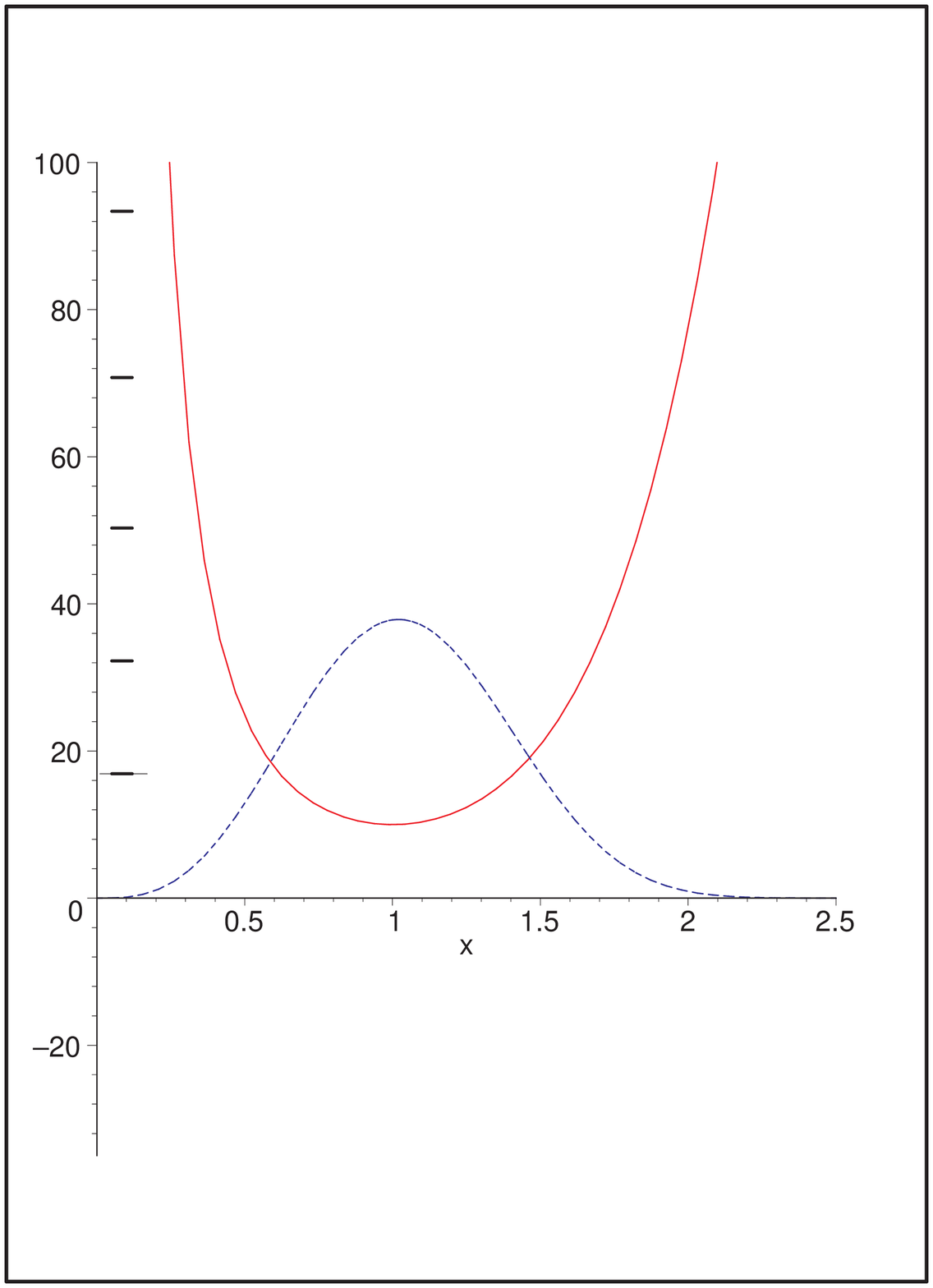}
\\[3pt]
\parbox{0.45\linewidth}{
\raggedright
{\small Figure \ref{p1}:
$\alpha=-9$, $l=0$, $E_2=16.919850$
}}
&
\parbox{0.5\linewidth}{
\raggedright
{\small Figure \ref{p2}:
$\alpha=3$, $l=2$, $E_0=16.919850$
}}
\end{array}
\]

\medskip

%
\[
\begin{array}{ll}
\refstepcounter{figure}
\label{pp1}
\includegraphics[width=0.460\linewidth,height=0.460\linewidth]{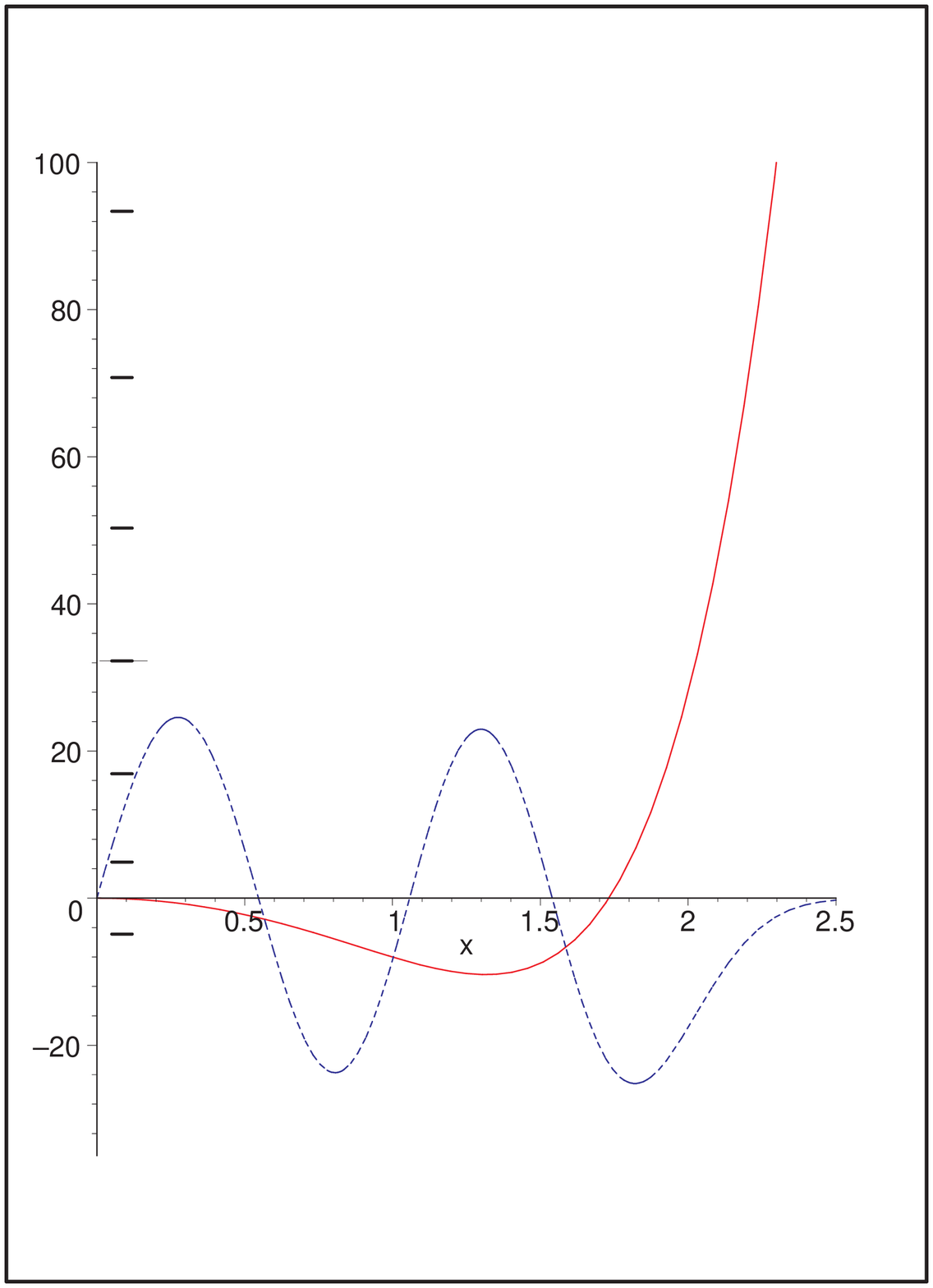}~~~~
&
\refstepcounter{figure}
\label{pp2}
\includegraphics[width=0.460\linewidth,height=0.460\linewidth]{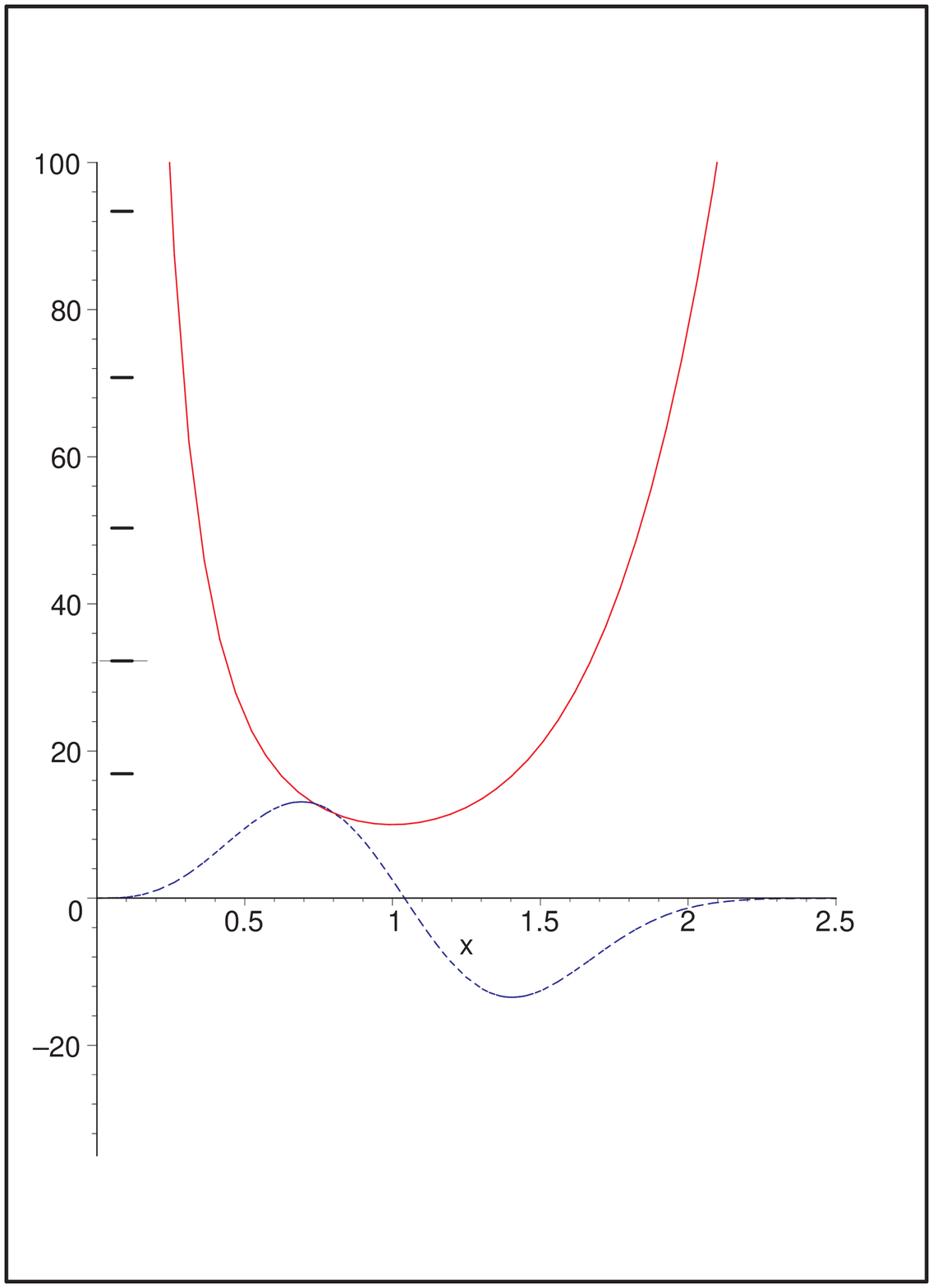}
\\[3pt]
\parbox{0.45\linewidth}{
\raggedright
{\small Figure \ref{pp1}:
$\alpha=-9$, $l=0$, $E_3=32.240265$
}}
&
\parbox{0.5\linewidth}{
\raggedright
{\small Figure \ref{pp2}:
$\alpha=3$, $l=2$, $E_1=32.240265$
}}
\end{array}
\]
%

%
\[
\begin{array}{ll}
\refstepcounter{figure}
\label{p3}
\includegraphics[width=0.460\linewidth,height=0.460\linewidth]{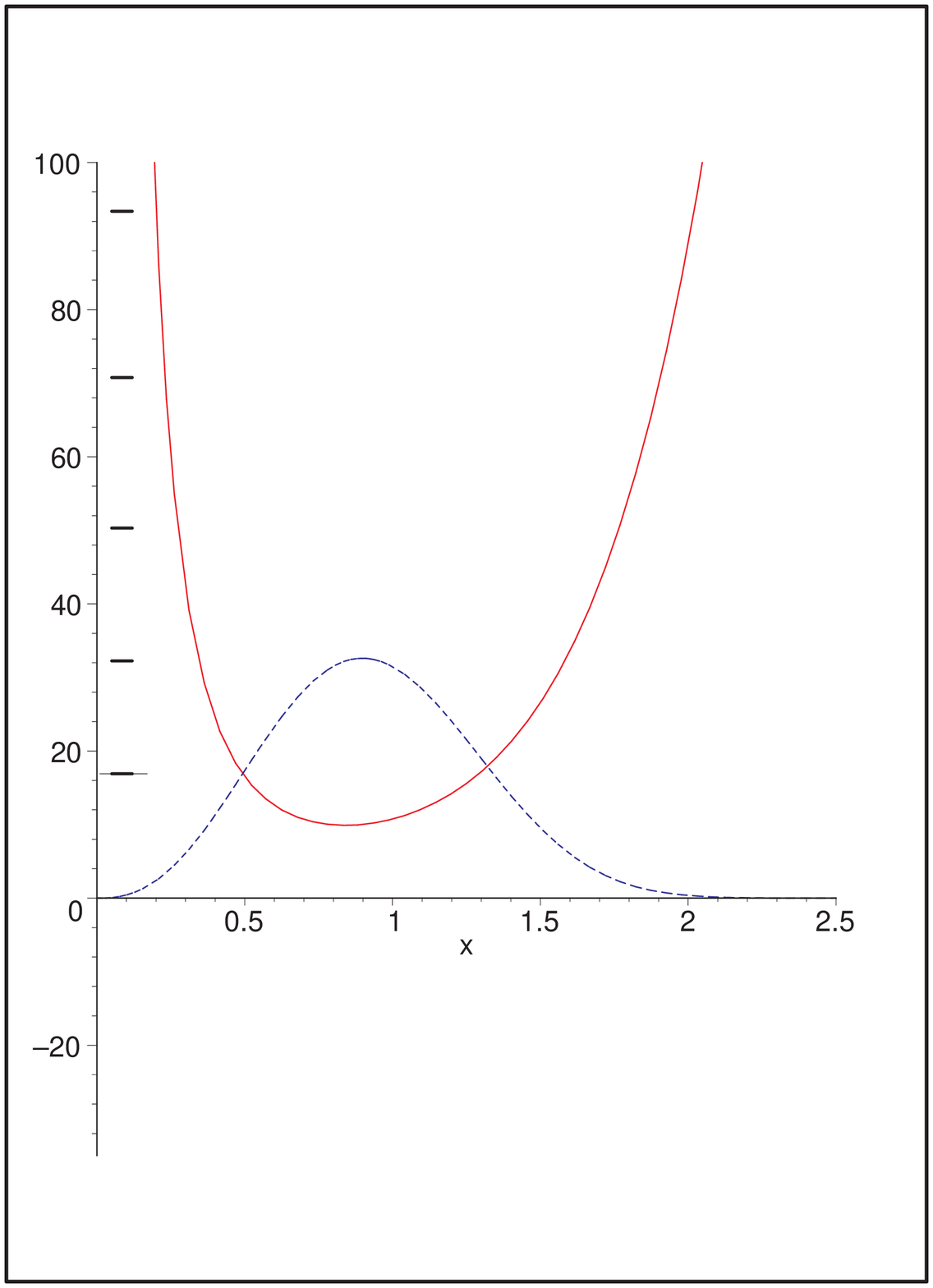}~~~~
&
\refstepcounter{figure}
\label{pp3}
\includegraphics[width=0.460\linewidth,height=0.460\linewidth]{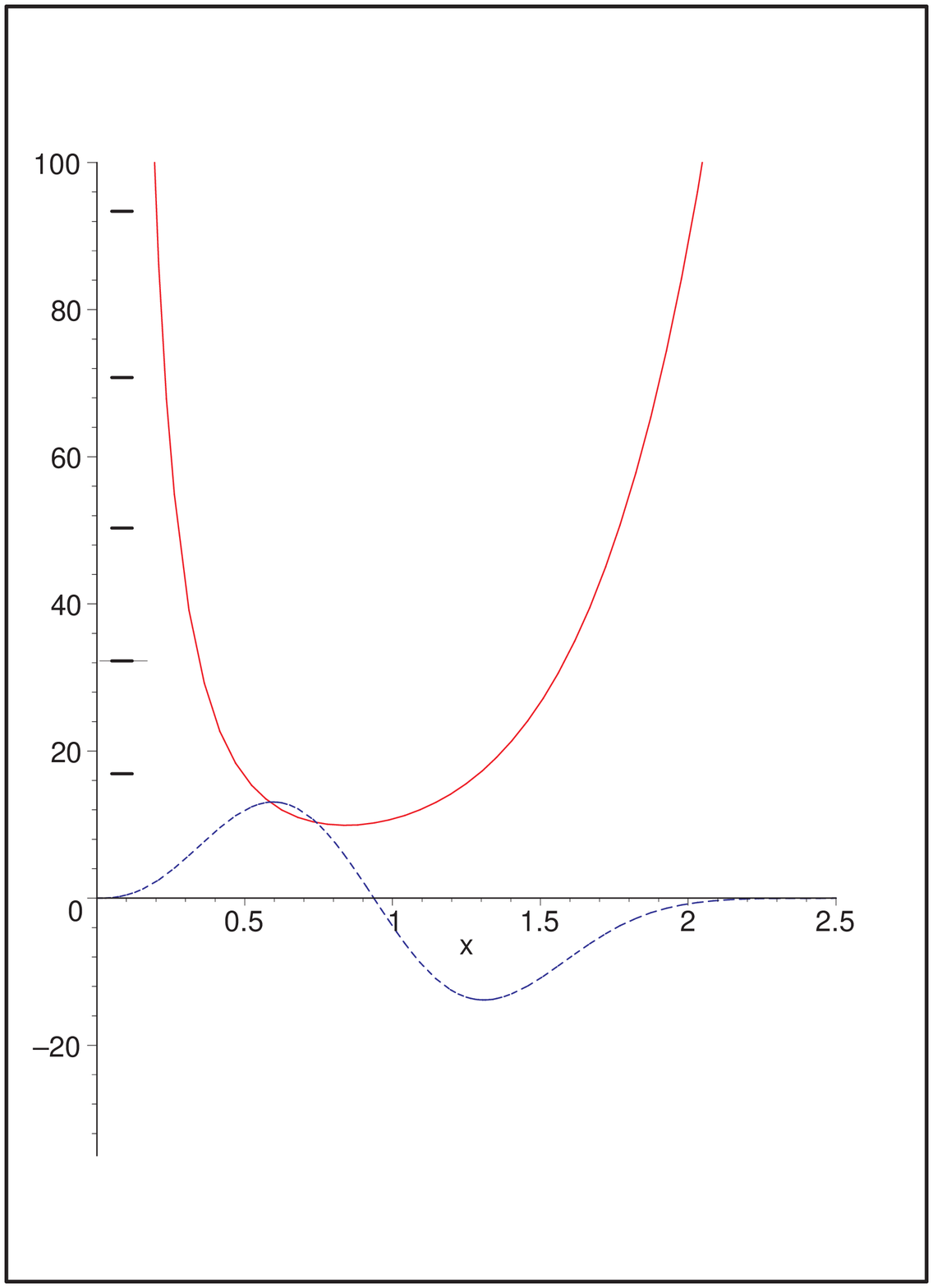}
\\[3pt]
\parbox{0.45\linewidth}{
\raggedright
{\small Figure \ref{p3}:
$\alpha=6$, $l=3/2$, $E_0=16.919850$
}}
&
\parbox{0.45\linewidth}{
\raggedright
{\small Figure \ref{pp3}:
$\alpha=6$, $l=3/2$, $E_1=32.240265$
}}
\end{array}
\]
%


\bigskip

Finally, in figures~\ref{p4} and \ref{pp4}
we illustrate how `extra' energy levels can appear with
the irregular boundary condition in resonance situations, a phenomenon that
was mentioned at the end of the conclusions above. 
Again, we take $J=2$.
To avoid numerical difficulties, in figure \ref{pp4} we
shifted the angular momentum slightly away from the exactly-resonant
value. 

\medskip

\[
\begin{array}{ll}
\refstepcounter{figure}
\label{p4}
\includegraphics[width=0.460\linewidth,height=0.460\linewidth]{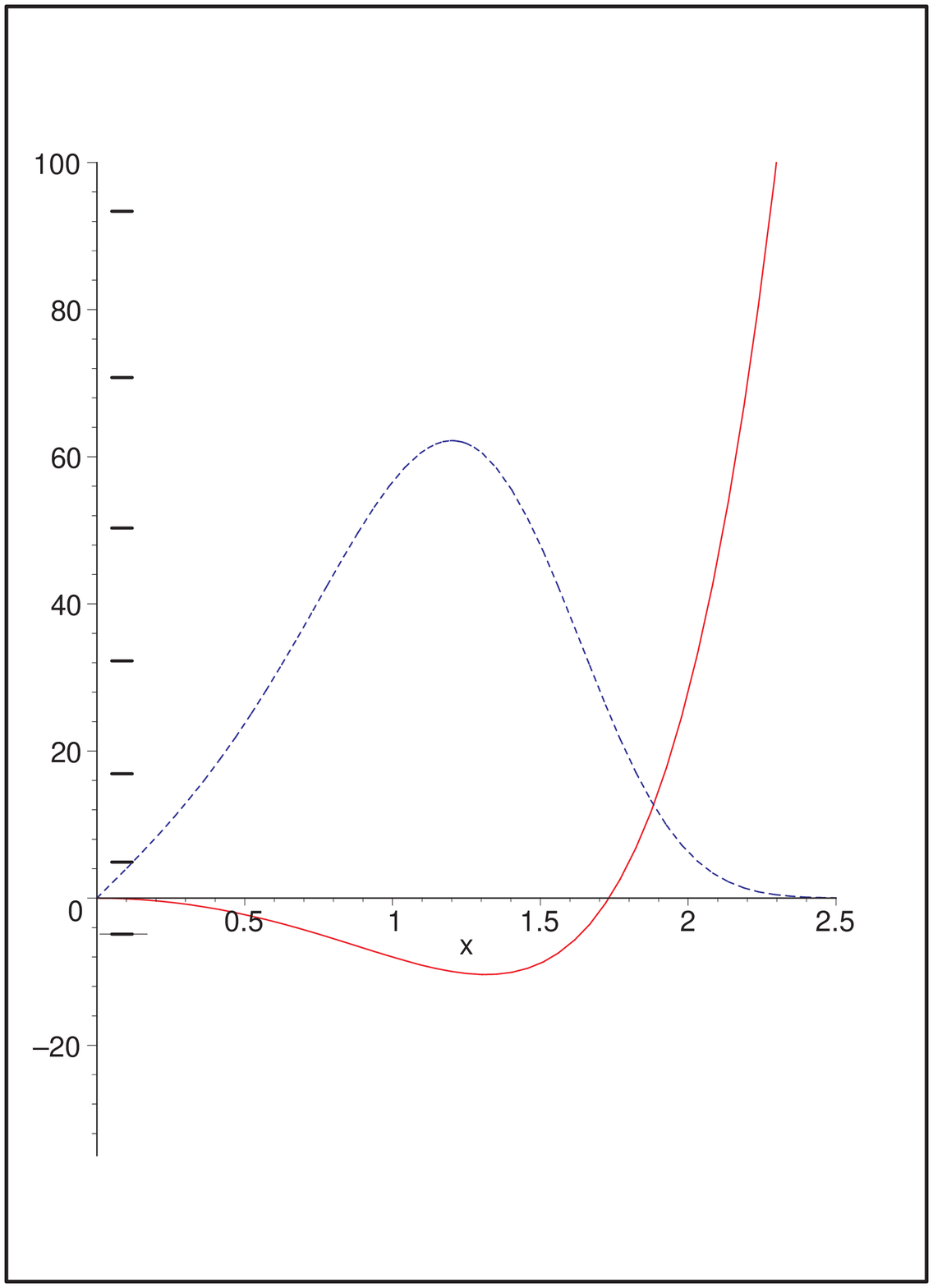}~~~~
&
\refstepcounter{figure}
\label{pp4}
\includegraphics[width=0.460\linewidth,height=0.460\linewidth]{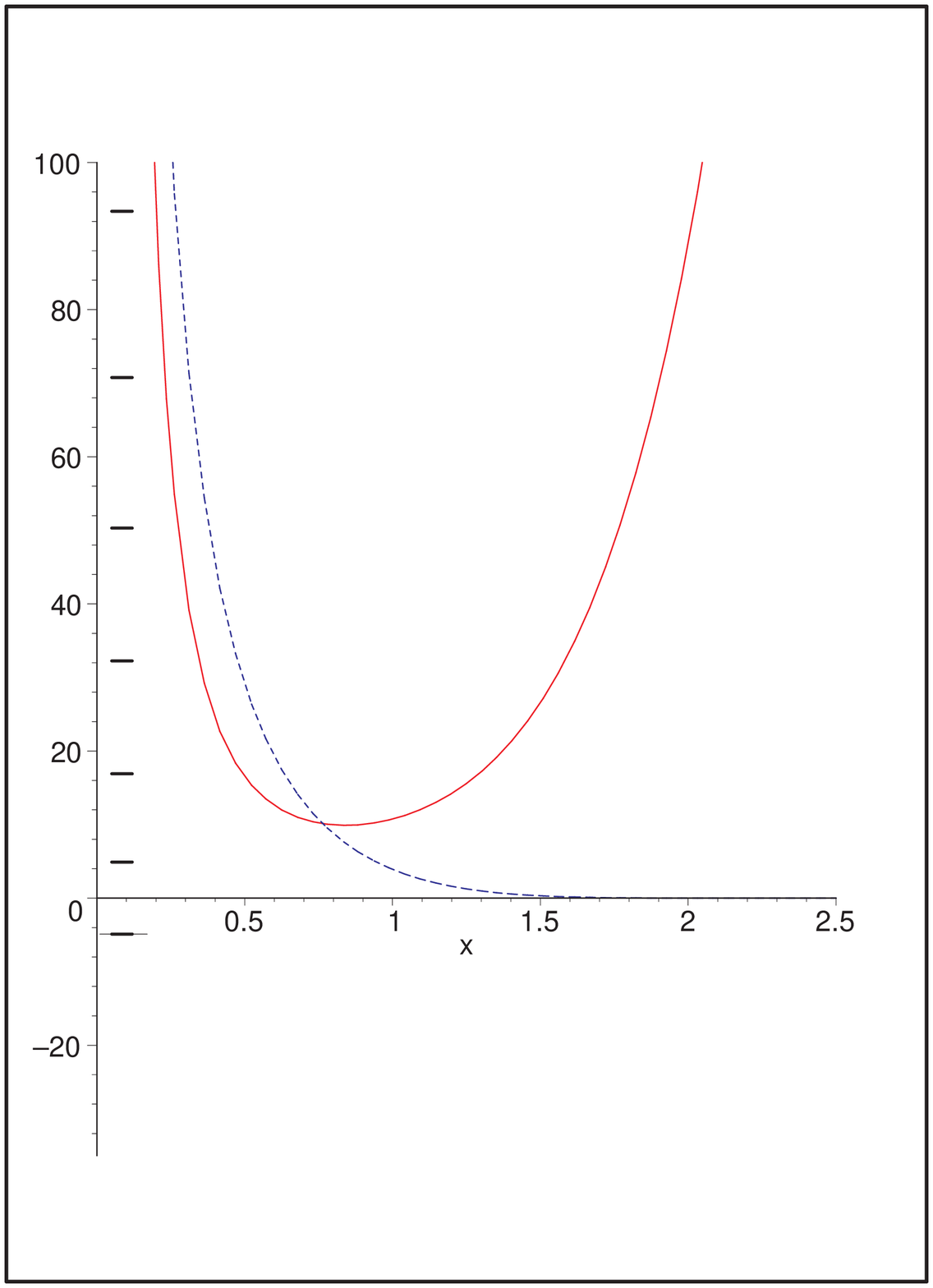}
\\[3pt]
\parbox{0.5\linewidth}{
\raggedright
{\small Figure \ref{p4}:
$\alpha=-9$, $l=0$, $E_0=-4.898979$}}
&
\parbox{0.5\linewidth}{
\raggedright
{\small Figure \ref{pp4}:
$\alpha=6$, $l=-5/2{+}10^{-6}$, $E_0=-4.898974$}}
\end{array}
\]

\smallskip

\resection{An elementary
proof of the Bessis, Zinn-Justin, Bender and Boettcher  conjecture}
\label{appbb}
A conjecture of Bessis and Zinn-Justin \cite{BZJ}, generalised by 
Bender and Boettcher \cite{BB}, 
states that the eigenvalues $\lambda_k$ of the
${\cal P T}$-symmetric Schr\"odinger equation
\eq
\Bigl[-\frac{d^2}{dx^2}-(ix)^{2M} \Bigr]\psi_k(x)=\lambda_k\,\psi_k(x)\, ,
\qquad\psi_k(x) \in L^2(\CaC)\, 
\label{PT}
\en
are real and positive for $M \ge 1$. 
The contour $\CaC$ on which the wavefunction is defined can be taken to be
the real axis for $M<2$\,; beyond this point, the
contour should be deformed down into the complex plane so as to remain in
the same pair of Stokes sectors \cite{BB}. 
(For an informal review in the context of the
ODE/IM correspondence, see also \cite{DDTr}.)
This conjecture has provoked a fair amount of work in recent years, a sample
being refs.~\cite{BBM,DP,DT,Mez,Zl,BCQ,BDMS,BW,Mez1}. 
In this appendix we consider
a slightly more general class of ${\cal PT}$-symmetric
spectral problems, namely
\eq
\Bigl[-\frac{d^2}{dx^2}-(i x)^{2M} 
-\alpha (i x)^{M-1}+ \frac{l(l+1)}{x^2}  \Bigr]\psi_k(x)=\lambda_k\,\psi_k(x)\, ,
{}~~~\psi_k(x) \in L^2(\CaC)\, 
\label{PTg}
\en
with $M$, $\alpha$ and $l$ real.
Again, for $M<2$
the contour $\CaC$ can be taken to be the real axis, though 
if $l(l{+}1)\neq 0$ it should be distorted so as to pass below 
the origin. 
We shall prove 
reality of the spectrum for $M>1$,
$\alpha<M{+}1{+}|2l{+}1|$, and positivity for $M>1$, $\alpha<M{+}1{-}|2l{+}1|$.
The spectrum might be real for a greater range of $\alpha$, but 
strict positivity certainly fails on the lines $\alpha=M{+}1{-}|2l{+}1|$.
Even with the restrictions on $\alpha$, our
result
includes the previously-considered cases:
for $\alpha=l(l{+}1)=0$ and $M=3/2$, a version of
the original Bessis -- Zinn-Justin conjecture is recovered; allowing $M$ to 
vary then gives the generalisation discussed by Bender and 
Boettcher, while the conjecture for
$\alpha=0$ and $l$ small 
was proposed in \cite{DTb}.
(Strictly speaking the original BZ-J conjecture concerned
the potential $x^2+igx^3$ with $g$ real; our discussion applies to the
strong-coupling limit of this problem.)

Setting $\Phi(x)= \psi(x/i)$, (\ref{PTg}) becomes 
\eq
\Bigl[-\frac{d^2}{dx^2}+x^{2M}+\alpha x^{M-1}+ 
\frac{l(l+1)}{x^2} \Bigr]\Phi_k(x)=-\lambda_k\,\Phi_k(x)\,,
{}~~~\Phi_k(x) \in L^2(i\CaC)\, ,
\label{shl}
\en
and has the same form as (\ref{sh}) with $E=-\lambda_k$, though with
different boundary
conditions: to qualify as an eigenfunction, $\Phi$
must decay as $|x|\to\infty$ along the contour $i\CaC$.
However, it is an easy generalisation of 
the discussion in \S7 of \cite{DTb} 
that the  function $T(-\lambda,\alpha,l)$ defined in 
(\ref{tform}) 
is the spectral determinant associated to the  
spectral problem (\ref{shl}). 
This identification allows us to study the
generalised BZ-JBB conjecture (\ref{PTg}) using techniques inspired by
the Bethe ansatz.

We start from   equation (\ref{tq}):  
\eq
T^{(+)}(E) D^{(+)}(E) =\Omega^{-(2l+1+\alpha)/2}  
D^{(-)}(\Omega^{2M} E)+
\Omega^{(2l+1+\alpha)/2} D^{(-)}(\Omega^{-2M}  E)\, ,
\label{tq11}
\quad
\en
and  define the zeroes of $T^{(+)}(E)=T(E,\alpha,l)$ to be the set 
$\{ -\lambda_k\}$. (Note that for $\alpha=0$, (\ref{tq11})
reduces to the T-Q system
obtained in \cite{DTb}.) Putting  
$E= -\lambda_k$ in (\ref{tq11}) and using, for $M>1$, the factorised 
form for  $D^{(-)}(E)$ gives the following constraints on the  $\lambda_k$'s:
\eq
\prod_{n=0}^{\infty} \lf( { E^{(-)}_n + \Omega^{-2M} \, \lambda_k  \over 
E^{(-)}_n +  \Omega^{2M} \, \lambda_k}
\ri)
= - \Omega^{-2l-1 - \alpha }\, ,\qquad k=0,1,\dots
\label{bb}
\en
Since the original eigenproblem (\ref{PTg}) is invariant under $l\to
-1{-}l$, we can assume $l\ge -1/2$ without any loss of generality.
Then each $E^{(-)}_n$ is an eigenvalue of
an Hermitian operator $\CH(M,-\alpha,l)$, and
hence is real. Furthermore a Langer transformation~\cite{La} 
(see also~\cite{BLZa,DTb})
shows that the
$E_n^{(-)}$ solve a
generalised eigenproblem with an
everywhere-positive `potential', and so are all 
positive, for $\alpha<1{+}2l$. This can be sharpened
by considering the value of $D^{(-)}(E)|\phup_{E=0}$\,. From
(\ref{Dzero}) below, this first vanishes when $\alpha=M{+}2l{+}2$. Until this
point is reached, no  eigenvalue $E^{(-)}_n$ can have passed the
origin, and all must be positive.
(It might be worried that negative eigenvalues could appear from
$E=-\infty$\,, but this possibility can be ruled out by a consideration
of the Langer-transformed version of the equation.)

Taking the modulus${}^2$ of (\ref{bb}), using 
the reality of the $E^{(-)}_k$, and writing the eigenvalues of
(\ref{PTg}) as $\lambda_k=|\lambda_k|\exp(i\,\delta_k)$, we have
\eq
\prod_{n=0}^{\infty} \lf ( { (E^{(-)}_n)^2 + |\lambda_k|^2  + 2  E^{(-)}_n
 |\lambda_k| \cos(\fract{2 \pi}{M+1} + \delta_k)
 \over
(E^{(-)}_n)^2 + |\lambda_k|^2  + 2  E^{(-)}_n
 |\lambda_k| \cos(\fract{2\pi}{M+1} - \delta_k)}
\ri)
= 1\, .
\label{abs}
\en
For $\alpha<M+2l+2$\,, all the $E^{(-)}_n$ are positive,
and each single term in the product on the LHS of (\ref{abs}) is either
greater than, smaller than, or equal to one 
depending  only on the relative values of the cosine terms in the numerator
and denominator. These are
independent of the index $n$.
Therefore the only possibility to match the RHS is for each term in the
product to be individually equal to one, which for $\lambda\phup_k\neq 0$
requires
\eq
\cos(\fract{2\pi}{M+1}+\delta_k)=\cos(\fract{2\pi}{M+1}-\delta_k) ~,
\quad
\mbox{or}
\quad
\sin(\fract{2\pi}{M{+}1})\sin(\delta_k)=0\,.
\en
Since  $M > 1$, this latter condition implies
\eq
\delta_k= n \pi\, ,~~~~n \in \ZZ 
\en 
and this establishes the reality of the
eigenvalues of (\ref{PTg})
 for $M>1$ and $\alpha<M+2l+2$ or, relaxing the condition on
$l$, $\alpha<M+1+|2l{+}1|$.

One might ask what goes wrong for $M<1$, since from \cite{BB} (and, for
the case $l\neq 0$, \cite{DTb}) it is
known that most of the $\lambda_k$ become complex as $M$ falls below $1$,
at least for $\alpha=0$.
The answer is
that if $M<1$, the order of $D^{(-)}(E)$ is greater than $1$,
the factorised form of $D^{(-)}(E)$ provided by Hadamard's theorem
no longer has such a simple form, and the proof breaks down. 

The borderline case
$M=1$ is the simple harmonic oscillator, exactly solvable for all 
$l$ and $\alpha$\,. Starting from the discussion in \S3 of \cite{DTb}, it is
easily seen that 
\eq
T(E,\alpha,l)|\phup_{M=1}=\frac{2\pi}
{\Gamma\Bigl(\frac{1}{2} +\frac{2l+1+E-\alpha}{4}\Bigr)
\Gamma\Bigl(\frac{1}{2} -\frac{2l+1-E+\alpha}{4}\Bigr)}
\en
and so the eigenvalues of (\ref{PTg}) are at
$\lambda=4n+2-\alpha\pm(2l{+}1)$, $n=0,1,\dots$~. All are real for
{\em all} real values of
$\alpha$ and $l$, and all are positive for $\alpha<2-|2l{+}1|$.

To discuss positivity at general values of $M>1$, we can continue in
$M$, $\alpha$ and $l$ away from a point in this latter region, $\{M{=}1,
\alpha<2-|2l{+}1|\}$.
So long as $\alpha$ remains less than $M+1+|2l{+}1|$, all eigenvalues will
be confined to the real axis during this 
process, and the first passage of an eigenvalue from positive
to negative values will be signalled by the presence of
a zero in $T(-\lambda,\alpha,l)$ at $\lambda=0$. 
Fortunately, $T(-\lambda,\alpha,l)|\phup_{\lambda=0}$ can be calculated exactly, 
extending an argument given for $\alpha=0$ in \cite{DTb}. First, one
notices that the function
\eq
\varphi(x)=\left(\fract{M{+}1}{2}\right)^{\frac{M{+}\alpha}{2M{+}2}}
x^{\frac{M{-}1}{2M{+}2}}\,Y\left(\left(\fract{M{+}1}{2}\right)^{\frac{1}{M{+}1}}
x^{\frac{2}{M{+}1}},E,\alpha,l\right)
\en
solves the Schr\"odinger equation
\eq
\Bigl[\,-\frac{d^2}{dx^2}+x^{2}-\sigma x^{\frac{2-2M}{M+1}}
+ \frac{\gamma(\gamma+1)}{x^2} 
\,\Bigr]
\varphi(x)=\Lambda \;\varphi(x) \, , 
\label{spectiii}
\en
where 
\eq
\sigma=\Bigl(\fract{2}{M{+}1}\Bigr)^{\frac{2M}{M+1}}E\,,~~~
\gamma=\frac{2l{+}1}{M{+}1}-\frac{1}{2}\,,~~~
\Lambda=-\frac{2\alpha}{M{+}1}~.
\en
(This transformation, which can be found via a pair of Langer
transformations, leads to equation (\ref{spectii}) in the case $M{=}3$.)
Further, $\varphi(x)$  has the large-$x$ asymptotic 
\eq
\varphi(x)\sim \frac{1}{\sqrt{2i}}\, x^{-\frac{1}{2}-\frac{\alpha}{M{+}1}}
\,\exp\left(-\frac{1}{2}x^2\right).
\en
At $E{=}0$, $\sigma=0$ and
(\ref{spectiii}) is the simple harmonic oscillator, which can be
solved exactly in terms of the confluent hypergeometric function $U(a,b,z)$. 
Matching asymptotics at large $x$, 
\eq
\varphi(x)|\phup_{E=0}=\frac{1}{\sqrt{2i}}\,x^{\gamma+1}\,e^{-x^2/2}
\,U\left(\,\fract{1}{2}(\gamma{+}\fract{3}{2}){-}\fract{1}{4}\Lambda,
\gamma{+}\fract{3}{2},x^2\,\right).
\en
Reversing the variable changes, extracting the leading behaviour as
$x\to 0$ and comparing with
(\ref{Ynear0}), we find
\eq
D(E,\alpha,l)|\phup_{E=0}=
D^{(+)}(E)|\phup_{E=0}=
\frac{1}{\sqrt{2i}}\Bigl(\fract{M{+}1}{2}\Bigr)^{\frac{2l+1-\alpha}{2M+2}
-\frac{1}{2}}
\frac{\Gamma\left(\frac{2l+1}{M+1}\right)}
{\Gamma\left(\frac{2l+1+\alpha}{2M+2}+\frac{1}{2}\right)}~.
\label{Dzero}
\en
Now $T(E,\alpha,l)|\phup_{E=0}$ follows from
(\ref{tq11}), remembering that $D^{(-)}(E)=D(E,-\alpha,l)$\,:
\eq
T(E,\alpha,l)|\phup_{E=0}=
T^{(+)}(E)|\phup_{E=0}=
\Bigl(\fract{M{+}1}{2}\Bigr)^{\frac{\alpha}{M+1}}
\frac{2\pi}
{\Gamma\left(\frac{1}{2} +\frac{2l+1-\alpha}{2M+2}\right)
\Gamma\left(\frac{1}{2} -\frac{2l+1+\alpha}{2M+2}
\right)}~.
\label{Tzero}
\en
The first zero arrives at $E=-\lambda=0$ when $\alpha=M+1-|2l{+}1|$, and so for
all $\alpha<M+1-|2l{+}1|$\,, the spectrum is entirely positive, as claimed.

In finishing, we return to the reality of the spectrum encoded by
$T(-\lambda)$. We have
proved that, if $M>1$, the eigenvalues $\lambda_k$ are real for all real
$\alpha<M+1+|2l{+}1|$.
One might conjecture that this reality should hold for {\em all} 
real $\alpha$ and $l$. However,
this is definitely not the case: for $M=3$, at the QES points and with $l$
sufficiently negative, an examination of the Bender-Dunne polynomials
shows that the exactly-calculable part of the spectrum of
$D(E,\balpha_J)$ has at least one pair of complex-conjugate
eigenvalues. The identity
$T(-E,\HH\balpha_J)=\gamma(\balpha_J)D(E,\balpha_J)$, which follows from the
results obtained in \S7 above, shows that 
$T(-E,\HH\balpha_J)$ must share these complex zeroes. If $(\alpha_J,l)$ are
real then so are $(\widehat\alpha_J,\widehat l_J)$, and so such examples
demonstrate that $T$ can have complex zeroes even while $M>1$ and $\alpha$ and
$l$ are real. It would be worthwhile to map out the full extent of the region
where the spectrum is entirely real, but we have not yet done this, beyond a
quick check that it appears to extend at least some way beyond the domain
$\alpha<M+1+|2l{+}1|$ covered by the proof given in this appendix.
%
%
%

\bigskip

{\small
\noindent {\it Note added in proof.}\\
(1) We have recently obtained some further results on the region within
  which the spectrum of the ${\cal P}{\cal T}$-symmetric problem discussed in
  appendix B becomes complex. These can be found in \cite{DDTs}.\\
(2) An alternative treatment of a class of ${\cal P}{\cal T}$-symmetric
 quantum mechanical problems similar to those discussed in appendix B can
 be found in \cite{Han,HKWT2}.
}


\bigskip

\begin{thebibliography}{99}
%
\raggedright
\parskip 1pt
%
\bibitem{Instref}
J. Zinn-Justin,
{\em Quantum Field Theory and Critical Phenomena},
(Clarendon, Oxford  1989) (International series of monographs on physics, 77)
%
\bibitem{Wit}
E. Witten,  
`Dynamical breaking  of  supersymmetry',
Nucl. Phys. B188 (1981) 513\toline{554}
%
\bibitem{Tur}
A.V. Turbiner, `Quasi-exactly-soluble problems and sl(2,R) algebra'
Comm. Math. Phys. 118 (1988) 467\toline{474}
%
\bibitem{Ush}
A.G. Ushveridze, {\em Quasi-Exactly Solvable Models in Quantum Mechanics},
(Institute of Physics, Bristol, 1993)
%
\bibitem{BD}
C.M. Bender and G.V. Dunne,  
{`Quasi-Exactly Solvable Systems and Orthogonal Polynomials'},
J. Math. Phys. 37 (1996) 6\toline{11},
{\tt hep-th/9511138  }
%
\bibitem{Lip}
L.N. Lipatov,
{`Duality symmetry of Reggeon interactions in multicolour QCD'}
Nucl. Phys. B548 (1999) 328\toline{362},
{\tt hep-ph/9812336 }
%
\bibitem{DTa}
P.\ Dorey and R.\ Tateo,
{`Anharmonic oscillators, the thermodynamic Bethe 
ansatz and nonlinear integral equations'},
J. Phys. A32 (1999) L419\toline{L425},
{\tt hep-th/9812211}
%
\bibitem{voros}
A.\ Voros,
{`Semi-classical correspondence and exact results: the case of the
spectra of homogeneous Schr\"odinger operators'}, 
J. Physique Lett. 43 (1982) L1\toline{L4};\\
{}~~---~~
{`The return of the quartic oscillator. The complex WKB method'},
Ann. Inst. Henri Poincar\'e Vol XXXIX (1983) 211\toline{338};\\
{}~~---~~
{`Exact resolution method for general 1D polynomial Schr\"odinger
equation'}, 
J. Phys. A32 (1999) 5993\toline{6007},
{\tt math-ph/9903045}; Corrigendum: J. Phys. A33 (2000) 5783\toline{5784}
%
\bibitem{BLZa}
V.V.\ Bazhanov, S.L.\ Lukyanov and A.B.\ Zamolodchikov, 
{`Spectral determinants for Schr\"odinger equation 
and Q-operators of Conformal Field Theory'},
J. Stat. Phys. 102
(2001) 567\toline{576},
{\tt hep-th/9812247}
%
\bibitem{Sa}
J.\ Suzuki,
{`Anharmonic Oscillators, Spectral Determinant and 
Short Exact Sequence of $U_q(\hat{sl}_2)$'},
J. Phys. A32 (1999) L183\toline{L188},
{\tt hep-th/9902053}
%
\bibitem{DTb}
P.\ Dorey and R.\ Tateo,
{`On the relation between Stokes multipliers and the
T-Q systems of conformal field theory'},
Nucl. Phys. B563 (1999) 573\toline{602},
{\tt hep-th/9906219}
%
\bibitem{DTc}
P.\ Dorey and R.\ Tateo,
{`Differential equations and integrable models: the $SU(3)$ case'},
Nucl. Phys. B571 (2000) 583\toline{606}, {\tt hep-th/9910102}
%
\bibitem{Sb}
J.\ Suzuki,
{`Functional relations in Stokes multipliers
and Solvable Models related to $ U_q(A_n^{(1)})$'},
J. Phys. A33 (2000) 3507\toline{3521},
{\tt hep-th/9910215}
%
\bibitem{Sc}
J.\ Suzuki,
{`Functional relations in Stokes multipliers -- 
Fun with $x^6+\alpha x^2$ potential'},
J. Stat. Phys. 102
(2001) 1029\toline{1047},
{\tt quant-ph/0003066}
%
\bibitem{DDT2}
P.\ Dorey, C.\ Dunning and R.\ Tateo,
{`Differential equations for general  $SU(n)$ Bethe ansatz systems'}, 
J. Phys. A33 (2000) 8427\toline{8442},
{\tt hep-th/0008039}
%
\bibitem{Suzatt}
J.\ Suzuki,
{`Stokes multipliers, Spectral Determinants and $T{-}Q$ relations'},
{\tt nlin-sys/0009006}
%
\bibitem{DDTr}
P.\ Dorey, C.\ Dunning and R.\ Tateo,
{`Ordinary differential equations and integrable models'},
JHEP Proceedings PRHEP-tmr 2000/034, {\em Nonperturbative Quantum
Effects 2000}, {\tt hep-th/0010148}
%
\bibitem{NEWT}
R.G.\ Newton,
{\it The complex $j$-plane}
(Benjamin 1964)
%
\bibitem{Sib}
Y.\ Sibuya,
{\it Global Theory of a second-order linear ordinary differential
operator with polynomial coefficient}\
(Amsterdam: North-Holland 1975)
%
\bibitem{BLZneq}
V.V.\ Bazhanov, S.L.\ Lukyanov and A.B.\ Zamolodchikov, 
`On non-equilibrium states in QFT model with boundary interaction',
Nucl. Phys. B549 (1999) 529\toline{545},
{\tt hep-th/9812247}
%
\bibitem{BG}
V.\ Buslaev and V.\ Grecchi,
`Equivalence of unstable anharmonic oscillators and double wells',
J. Phys. A26 (1993) 5541\toline{5549}
%
\bibitem{BZJ}
D.\ Bessis and J.\ Zinn-Justin,
unpublished
%
\bibitem{BB}
C.M.\ Bender and S.\ Boettcher,
{`Real spectra in non-hermitian Hamiltonians having $\cal{PT}$ symmetry'},
Phys. Rev. Lett. 80 (1998) 4243\toline{5246},
{\tt physics/9712001}
%
\bibitem{BBM}
C.M.\ Bender, S.\ Boettcher and P.N.\ Meisinger,
{`$\cal{PT}$-symmetric quantum mechanics'},
J. Math. Phys. 40 (1999) 2201\toline{2229},
{\tt hep-th/9809072}
%
\bibitem{DP}
E.\ Delabaere and F.\ Pham,
`Eigenvalues of complex Hamiltonians with ${\cal PT}$-symmetry. I',
Phys. Lett. A250 (1998) 25\toline{28}
%
\bibitem{DT}
E.\ Delabaere and D.\ Trinh,
`Spectral analysis of the complex cubic oscillator',
J. Phys. A33 (2000) 8771\toline{8796}
%
\bibitem{Mez}
G.A.\ Mezincescu,
`Some properties of eigenvalues and eigenfunctions of the cubic oscillator
with imaginary coupling constant',
J. Phys. A33 (2000) 4911\toline{4916},
{\tt quant-ph/0002056}
%
\bibitem{Zl}
M.\ Znojil,
`Spiked and PT-symmetric decadic potentials supporting elementary
$N$-plets of bound states',
J. Phys. A33 (2000) 6825\toline{6833},
{\tt quant-ph/0002083}
%
\bibitem{BCQ}
B.\ Bagchi, F.\ Cannata and C.\ Quesne,
`PT-symmetric sextic potentials', 
Phys. Lett. A269 (2000) 79\toline{82},
{\tt quant-ph/0003085}
%
\bibitem{BDMS}
C.M.\ Bender, G.V.\ Dunne, P.N.\ Meisinger and M.\ Simsek,
`Quantum complex H\'enon-Heiles potentials',
{\tt quant-ph/0101095}
%
\bibitem{BW}
C.M.\ Bender and Q.\ Wang,
`Comment on a recent paper by Mezincescu',
J. Phys. A34 (2001) 3325\toline{3328}
%
\bibitem{Mez1}
G.A.\ Mezincescu,
`The operator $p^2 -(i x)^{\nu}$ on $L^2(R)$ (reply to Comment by
Bender and Wang)', 
J. Phys. A34 (2001) 3329\toline{3332}
%
\bibitem{nfold}
H.\ Aoyama, M.\ Sato and T.\ Tanaka,
`General forms of a ${\cal N}$-fold supersymmetric family',
Phys. Lett. B503 (2001) 423\toline{429},
{\tt quant-ph/0012065}
%
\bibitem{AIS}
A.A.\ Andrianov, M.V.\ Ioffe and V.P.\ Spiridonov,
`Higher-derivative supersymmetry and the Witten index',
Phys. Lett. A174 (1993) 273\toline{279}, {\tt hep-th/9303005}
%
\bibitem{nfoldo}
H.\ Aoyama, H.\ Kikuchi, I.\ Okouchi, M.\ Sato and S.\ Wada,
`Valley views: instantons, large order behaviors, and supersymmetry',
Nucl. Phys. B553 (1999) 644\toline{710},
{\tt hep-th/9808034}
%
\bibitem{KP}
S.M.\ Klishevich and M.S.\ Plyushchay,
`Nonlinear supersymmetry, quantum anomaly and quasi-exactly solvable
systems',
Nucl. Phys. B606 (2001) 583\toline{612}
{\tt hep-th/0012023}
%
\bibitem{LS}
W.\ Lucha and F.F.\ Sch\"oberl,
`Solving the Schr\"odinger equation for bound states with Mathematica 3.0',
Int. J. Mod. Phys. C10 (1999) 607\toline{620},
{\tt hep-ph/9812368}
%
\bibitem{cheng}
H.\ Cheng,
`Meromorphic property of the $S$ matrix in the complex plane of angular momentum',
Phys. Rev. 127 (1962) 647\toline{648}
%
\bibitem{La}
R.E.\ Langer,
`On the connection formulas and the solutions
 of the wave equation',
Phys. Rev. 51 (1937) 669\toline{676}
%
\bibitem{DDTs}
P.\ Dorey, C.\ Dunning and R.\ Tateo,
`Supersymmetry and
 the spontaneous breakdown of ${\cal P}{\cal T}$-symmetry',
J. Phys. A34 (2001) L391\toline{L400}, 
 {\tt hep-th/0104119}
%
\bibitem{Han}
 C.\ R.\ Handy, 
`Generating converging bounds to the (complex) discrete
 states of the $P^2 + iX^3 + iX$ Hamiltonian', 
J. Phys.  A34 (2001) 5065\toline{5081}, 
{\tt math-ph/0104036}
%
\bibitem{HKWT2}
 C.\ R.\ Handy, D.\ Khan, Xiao-Qian Wang and C.\ J.\ Tymczak, 
`Multiscale reference function analysis of the ${\cal P}{\cal T}$
 symmetry breaking solutions for the $P^2+iX^3+i\alpha X$ Hamiltonian', 
J. Phys. A34 (2001) 5593\toline{5602}, 
{\tt  math-ph/010403}
%
\end{thebibliography}
\end{document}